\documentclass[twocolumn]{openjournal}

\usepackage[english]{babel}
\usepackage{ucs}
\usepackage[utf8x]{inputenc}
\usepackage[T1]{fontenc}

\usepackage{graphicx} 
\usepackage[colorlinks=False]{hyperref} 
\usepackage{amssymb}  %

\newcommand\colzero {\null}
\newcommand\cola {\null}
\newcommand\colb {&}
\newcommand\colc {&}
\newcommand\cold {&}
\newcommand\cole {&}
\newcommand\colf {&}
\newcommand\colg {&}
\newcommand\colh {&}
\newcommand\coli {&}
\newcommand\colj {&}
\newcommand\colk {&}
\newcommand\coll {&}
\newcommand\eol{\\}
\newcommand\extline{&&&&&&&&&&&\eol}

\begin{document}


\title{Star formation in the high-extinction Planck cold clump PGCC~G120.69+2.66}



\author{Anlaug Amanda Djupvik}
\affiliation{
Nordic Optical Telescope, Rambla Jos\'{e} Ana Fern\'{a}ndez P\'{e}rez 7, ES-38711 Bre\~{n}a Baja, Spain}
\affiliation{Department of Physics and Astronomy, Aarhus University, Munkegade 120, DK-8000 Aarhus C, Denmark} 

\author{Jo\~ao L. Yun}
\affiliation{Instituto de Astrof\'{\i}sica e Ci\^encias do Espa\c co (IA), Universidade de Lisboa \\   Departamento de Física, Faculdade de Ciências, Campo Grande, PT-1749-016 Lisboa, Portugal}
\affiliation{Visiting astronomer, Departament de Física Quàntica i Astrofísica, Institut de Cièncias del Cosmos, Universitat de Barcelona (ICC-UB), Martí i Franquès 1, E-08028 Barcelona, Spain}
    
\author{Fernando Comer\'{o}n}
\affiliation{European Southern Observatory, Karl-Schwarzschild-Str. 2, 85748 Garching bei München, Germany}

\begin{abstract}
   We investigate the star formation occurring in the Planck Galactic cold clump PGCC 120.69+2.66.
   Near-infrared JHK$_S$ images and K-band spectroscopy obtained with NOTCam at the Nordic Optical
    Telescope complemented with archive data are used to study the stellar content. In addition, millimetre
    line CO and CS spectra were obtained with the Onsala 20~m telescope, and sub-millimetre continuum SCUBA-2
    archive data are used to characterise the host molecular cloud.
    We identify a molecular cloud core traced by CO and CS emission at a distance of 1.1 kpc. In the region
    studied, we identify 5 submm continuum cores. Embedded in and around these dense submm cores, we find
    38 young stellar objects, classified as 9 Class~I, 8 Class~II, and 21 near-IR excess or variability
    sources, accompanied by bipolar nebulosities and signs of
    protostellar jets. Furthermore, a very bright and reddened source is found towards this molecular
    cloud core. Even though its location appears to suggest its association to the star formation region,
    its infrared spectral type is compatible with a red supergiant, hidden behind 36~mag of visual extinction.
    \end{abstract}

\section{Introduction}

Detecting and classifying the young stellar population in molecular clouds at an early star formation
phase is vital for the understanding of star formation process - the young stellar objects (YSOs) are
the end products and the molecular cloud cores are the initial conditions in which they form. Stars
tend to form in groups, often in rich clusters with hundreds to thousands of stars, but also in looser
aggregates with fewer young stellar objects \citep[][]{strom93}.

The presence of embedded jets and of extended nebular emission often reveals the young stellar
character of the associated sources \citep[e.g.,][]{mccaughrean94,yun97,yun2001}.
Recent large scale studies of nearby star formation regions with infrared surveys
\citep[e.g.][and references therein]{evans2009,gutermuth2009,winston2020} have investigated large
populations of embedded YSOs in their early evolutionary phases at high sensitivity and spatial
resolution, locating their birthplace to the densest molecular cloud cores, often found distributed
along filaments as traced by submm surveys \citep{andre2014}.  

Low-mass YSOs are commonly classified following their evolutionary stages \citep{adams87,andre93} {\it i)} as Class~0 (protostars heavily embedded in their parent molecular cloud cores only seen in the submm wavelengths); {\it ii)} as Class~I (infrared protostars surrounded by accreting envelopes and circumstellar discs); {\it iii)} as Class~II (optical visible sources displaying near-infrared excesses due to the presence of their circumstellar discs); and finally {\it iv)} as  transitional objects, pre-main-sequence stars close to reaching the main-sequence.

Based on previous investigations of star formation in the outer Galaxy, selecting IRAS
sources with colours and temperature maps strongly correlated with the presence of embedded infrared
protostars or young clusters \citep[e.g.][]{yun2001,palmeirim2010,yun2015}, we found that the region around
the $IRAS$ source {IRAS~00267+6511} was a promising candidate for an active star formation site
towards the outer Galaxy (l = 120.7$^{\circ}$, b = 2.7$^{\circ}$). The
region has been identified as a dense cloud core via CO, HCO+, and HCN emission \citep{zhang2016,yuan2016}
and is listed in the Planck catalog of Galactic Cold Clumps \citep{planck2016} derived from three
high-frequency bands (353, 545, and 857 GHz). The stellar content had not been studied, given that
at optical wavelengths the region is a patch of obscuration against background stars with no sign of
star formation. 
Candidate YSOs had been proposed around {IRAS~00267+6511} via application of automatic algorithms (e.g.\ \citet{wilson2023} use a naive Bayes classifier for identifying candidate Class II YSOs in the GAIA DR3, or via vector machine selection of WISE sources \citep[e.g.][]{marton2016}).  Here, we seek to determine the YSO population down to fainter magnitudes and with better angular resolution than presently available for this region.

Molecular cloud cores are known to exhibit very high values of visual extinction
hiding from optical view both YSOs forming in the cores and other background objects, as well. In fact,
in the 2MASS \citep{skrutskie2006} near-IR images of the region, we became aware of an intriguingly bright
and red object that caught our attention; invisible in optical surveys and situated only 34$\arcsec$ from
the IRAS source, i.e. close to the center of the cloud, it could be a very bright YSO or instead a
background unrelated source.

We obtained high-resolution near-infrared JHK$_S$ images and K-band spectroscopy with the Nordic Optical
Telescope \citep{djup2010} to identify and characterise young stellar sources. In addition, we use
millimetre and sub-millimetre data, both new and archive data, to define the molecular and dust
environment. 
We report here evidence for early star formation activity in the form of nebular emission, mid-- and
near-infrared excess sources in the protostellar and pre-main sequence phase. The bright red source
appears to be a previously unknown background massive supergiant star hidden behind several tens of
magnitudes of visual extinction. Section~\ref{obs} describes the observations and data reduction, in
Sects.~\ref{results} and~\ref{discussion} we present and discuss the results, and a summary is given
in Sect.~\ref{summary}.

\section{Observations}
\label{obs}
\subsection{Near-infrared observations}


\smallskip
Near-IR imaging and spectroscopy was obtained with NOTCam, the Nordic Optical Telescope's
near-IR Camera and Spectrograph\footnote{NOTCam is documented in detail at
\url{https://www.not.iac.es/instruments/notcam/}.} The NOTCam detector is a Hawaii 1k
HgCdTe IR array and the camera optics has two different scales \citep{abbo2000}.
The wide-field (WF) camera has pixel size of $0'.{234}$ and a $4\arcmin \times$
4$\arcmin$ field-of-view, and the high-resolution (HR) camera has a pixel size of
$0'.{079}$ and a fov of about 80$\arcsec \times$ 80$\arcsec$.

NOTCam broad-band JHK$_S$ imaging and narrow-band imaging on and off
the H$_2$ v=1-0~S(1) line at 2.122 $\mu$m, using filters \#218 and \#230, were obtained
in free slots in service nights with starting dates 2023-11-23, 2023-12-28 and 2024-07-17. 
 For the K$_S$ band we used a 6~mm cold stop to avoid saturation on the bright
  star. (Since NOTCam
  is equipped with cold stops of different diameters, we could reduce the 15~mm beam inside the
  instrument, effectively reducing the telescope size and lowering the transmission to 15\%.)
The total on-source exposure times used with the WF camera were 180s (J), 194s (H), 32s
(K$_S$ + 6mm stop), and 2300s (\#218 and \#230). Two smaller fields mapped with the
HR-camera, one to the North (N) and one to the North-West (NW) of the near-IR luminous star,
avoiding the bright source, and the total on-source exposure times were 540s in each filter J, H,
and Ks for both of these fields. The seeing measured in these images have PSF fwhm of
0.4$\arcsec$ and 0.6$\arcsec$ for the N and the NW fields, respectively.

Differential twilight flats were obtained for all filters and domeflats for the combination
of K-band and the 6mm stop. Deeper images were obtained with the HR-camera in the JHKs filters
in smaller regions near the bright source. The imaging was obtained with small-step dithers,
using ramp-sampling readout and individual exposure times short enough to avoid saturation.
The images were reduced with IRAF\footnote{IRAF is currently supported and maintained by
an IRAF community at https://iraf.net/.}
and the IRAF package notcam.cl v2.6, made for NOTCam
image reductions, including bad-pixel correction, flat-field correction, sky evaluation and
subtraction, distortion correction, shifting and combination the individual frames.

For each field and filter point sources are detected and measured with aperture photometry,
using the IRAF photometry package, applying aperture corrections found using curve of growth
on a few bright and isolated stars in each image. We cross-correlated each field with 2MASS
sources with quality flag AAA, used to calibrate the NOTCam JHK$_S$ photometry, as well
as the astrometry. The positional rms values are 0.05$\arcsec$ and 0.02$\arcsec$ for
the WF camera and the HR-camera, respectively. The photometric calibration errors of 0.021,
0.027, and 0.054 mag for J, H, and K$_S$, respectively, were added in quadrature to the
photometric errors obtained in the WF field. For the small HR fields, calibration sources
were limited to a few stars only. In the two small fields observed on different nights, there
is a star that differ by less than 0.04 mag in the H and K$_S$ bands and by 0.12 in J, the latter
most likely due to the faintness in the J-band (18 mag), and we decided to apply 0.04 mag as
calibration error for the photometry in the HR fields.

The spectroscopy was done with the NOTCam echelle grism Gr\#1, the K-band filter (\#208) used
as an order sorter, and the 128 $\mu$m wide slit (corresponding to $0.6\arcsec$ or 2.6 pixels).
This setup covers the range from 1.95 to 2.37 $\mu$m with a dispersion of 4.1 \AA /pix,
giving a spectral resolution $\lambda / \Delta \lambda$ of about 2100. K-band spectroscopy of
the bright source was obtained on 2023-11-08, dithering along two positions on the slit
(A-B) three times to obtain 6 individual spectra, each with an exposure time of 20 seconds
(reading out every 4 seconds 5 times in ramp-sampling mode). Calibration lamps were taken
while pointing to target to minimize flexure in the wavelength calibration and improve
fringe corrections with the spectroscopic flats. To correct the spectrum for Earth's
atmospheric features we used as a telluric standard the star HD5071 (SpT B8, $K_{\rm 2MASS}$ =
7.49 mag), observed right after the target. Hot-pixel masks were constructed from darks
taken with the same integration time and readout mode. For the target and the standard the
individual ABABAB exposures were corrected for hot pixels, flat-field corrected using
halogen flats, and sky-subtracted using the neighbouring frame. The individual 1D spectra
were optimally extracted and wavelength calibrated using dedicated scripts based on IRAF
tasks, after which they were combined to a final spectrum. The target spectrum was divided
by the telluric spectrum using the interactive IRAF task {\em telluric}, having first
interpolated over the stellar Br$\gamma$ absorption line at 2.166~$\mu$m in the telluric.
Differential photometry of the telluric in the acquisition image, using three reliable 2MASS
stars, gave $K$ = 7.56 $\pm$ 0.02~mag, differing by 0.07 mag from the published 2MASS value.
This calibration is used to flux-scale a Vega continuum which was multiplied with the target
spectrum to correct the slope and give a rough estimate of the target flux.

\subsection{CO and CS mm line observations}

Millimetre single-dish observations of IRAS~00267+6511 were carried out at the 20~m  Onsala
telescope, in 2009~April. Three maps were obtained in the rotational lines of $^{12}$CO(1$-$0)
(115.271~GHz), $^{13}$CO(1$-$0) (110.201~GHz), and CS(2$-$1) (97.981~GHz), respectively. 
The telescope half-power beam width (HPBW) was $32\arcsec$ at 115~GHz 
and $38\arcsec$ at 98~GHz. The adopted grid spacing was 30$''$ (approximately full-beam
sampling, adopted due to the limited observing time available). The $^{12}$CO(1$-$0),  $^{13}$CO(1$-$0), and CS(2$-$1) maps are composed of
$5 \times 5$, of $7 \times 7$, and of $5\times 6$ pointings, respectively.

A high-resolution 1600 channel spectrometer was used as a back end, with a total bandwidth
of 40~MHz and a channel width of 25~kHz, corresponding to approximately $0.065-0.075$~km~s$^{-1}$,
at those frequencies. The spectra were taken in frequency-switching mode, recommended to save
observational time when mapping extended sources. The antenna temperature was calibrated with
the standard chopper wheel method. Pointing was checked regularly towards known circumstellar
SiO masers; pointing accuracy was estimated to be better than $4\arcsec$.

Data reduction followed standard steps: \textit{i)}  folding the frequency-switched spectrum; 
\textit{ii)} fitting the baseline by a polynomial and subtracting it; \textit{iii)} coadding
repeated spectra obtained at the same sky position; \textit{iv)} obtaining the main beam
temperature $T_{\mathrm{MB}}$ by dividing the antenna temperature $T_{\mathrm{A}}$ by the
$\eta_{\mathrm{MB}}$ factor, equal to about 0.5 for these lines; and \textit{v)} finally, smoothing
the spectra to a velocity resolution of approximately $0.13-0.15$~km~s$^{-1}$ at those frequencies.
The spectrum baseline RMS noise (in $T_{\mathrm{MB}}$), averaged over all map positions, has been
found to be 1.4~K for $^{12}$CO(1$-$0), 0.4~K for $^{13}$CO(1$-$0), and 0.2~K for CS(2$-$1).

\subsection{Archive data}
 
This region was surveyed by the Herschel satellite whose database is accessible via the NASA/IPAC
Infrared Science Archive (IRSA). It was also observed by the Planck satellite. In this region, the 
{\it Planck Early Release Compact Source Catalogue} \citep{planck2011} lists a submm clump named
PLCKECC~G120.67+2.66. This catalog was later superseded by the {\it Planck Catalogue of Galactic
Cold Clumps} \citep{planck2016} containing the source PGCC~G120.69+2.66 (with slightly different
coordinates). This clump extends by $11'.6 \times 4'.4$ on the sky. The centre is positioned about
2 arcmin southeast of IRAS~00267+6511, and involves the whole region imaged in this study.

This region was covered by the SCUBA-2 Continuum Observations of Pre-protostellar Evolution Large
Program (SCOPE) survey 
\citep{liu2018,eden2019}.
The survey was conducted at 850~$\mu$m with a
beam size of 14.4 arcsec on the James Clerk Maxwell Telescope
equipped with the SCUBA-2 instrument \citep{holland2013}. We accessed the data base via the Canadian
Astronomy Data Centre and downloaded the corresponding FITS image of this region. With a pixel
scale of $4''$~pix$^{-1}$, it covers about $10' \times 10'$ encompassing the region mapped by
NOTCam and the CO and CS observations. 

\begin{figure}
  \includegraphics[width=9cm]{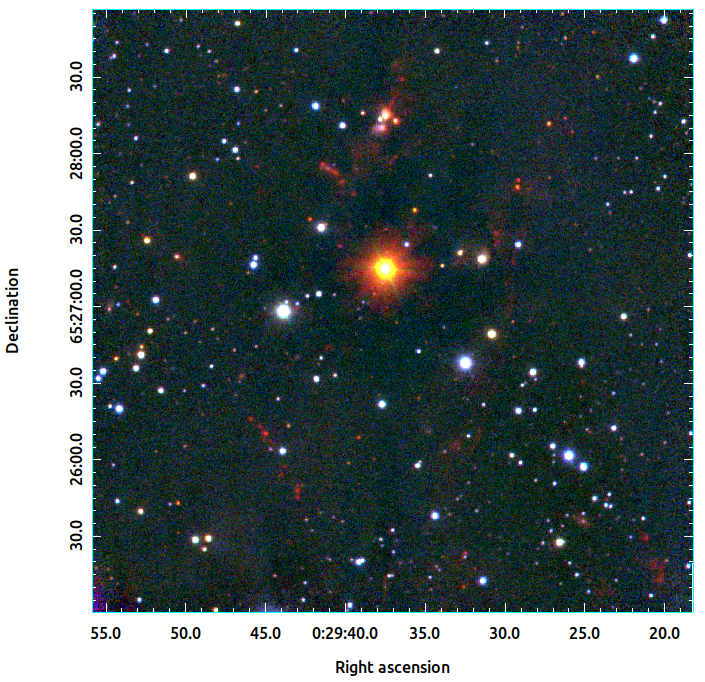}
  \caption{NOTCam RGB image of the Planck cold clump PGCC~G120.69+2.66 region using J-band
    (blue), H-band (green) and 2.122 $\mu$m H$_2$ line (red).
    The bright, red star near the centre is discussed in Sect.~\ref{sg-results}. }
\label{notcam-wf}
\end{figure}

\section{Results}
\label{results}

\subsection{The mid-infrared view}
\label{mir-results}

\begin{figure}
  \includegraphics[width=8.5cm]{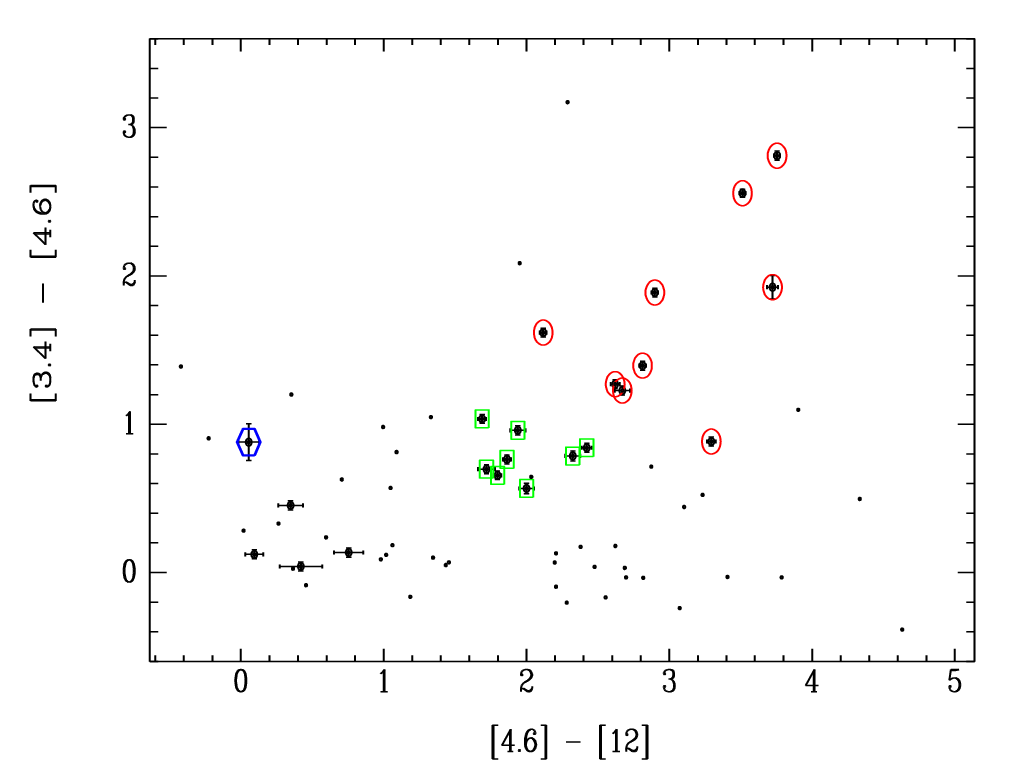}
  \caption{WISE [3.4]-[4.6]/[4.6]-[12] colour-colour diagramme of all 68 sources in the
      4' $\times$ 4' FOV. The 22 sources with quality flag A in all the three filters are shown as
      larger dots with error bars. The classified Class~I sources (red circles) and the Class~II
    sources (green squares) are shown. The blue hexagon
    marks the position of the brightest near-IR source in the region, a candidate red
    supergiant, see Sect.~\ref{sg-results}.}
  \label{wise-cc}
\end{figure}

We have explored the WISE \citep{wright2010} and 2MASS \citep{skrutskie2006} catalogues in the
4$\arcmin$ $\times$ 4$\arcmin$ region around IRAS~00267+6511 shown in the NOTCam image in
Fig.~\ref{notcam-wf}. After weeding out extragalactic contaminants and low S/N sources in
the WISE sample, we search for Young Stellar Objects (YSOs) among the sources based on their
mid-IR Spectral Energy Distributions (SEDs). The standard classification scheme for YSOs is
based on the shape of the mid-IR SED quantified by the SED index
$\alpha = d \log \lambda f_{\lambda}/d \log \lambda$ originally taken from 2 to 10 (or 25)
$\mu$m \citep{lada1984,greene1994}. A steeply rising SED towards longer wavelengths for
protostars embedded in dust envelopes, Class~I sources ($\alpha > 0.3$), can be easily
separated from a declining slope of the more evolved Class~II sources ($-1.6 < \alpha \le
-0.3$) surrounded by circumstellar dust disks. An intermediate range would be occupied by
the so-called Flat-Spectrum sources ($-0.3 < \alpha \le 0.3$). The SED index is frequently
translated to colour indices in various magnitude systems that will define IR excess
emission above the Rayleigh-Jeans tail of the photospheric black-body. For the WISE
photometry in the bands W1 (3.4 $\mu$m), W2 (4.6 $\mu$m) and W3 (11.8 $\mu$m) we apply
the criteria described by \citet{koenig2014} for removing extragalactic contaminants as well
as defining the Class~I and Class~II sources according to their location in the
$[3.4]-[4.6]/[4.6]-[12]$ colour-colour diagramme. Following their criteria we discard 75\% of
the sources in the field as having no mid-IR excess. These are stars, extragalactic
  sources, or low S/N detections. Only 22 of the 68 sources have photometric
  quality flag A in all three filters.
We find 17 YSOs with mid-IR excess and classify them as Class~I and Class~II sources as
shown in Fig.~\ref{wise-cc}, where the 9 Class~I sources are marked with red circles while
the 8 Class~IIs are marked with green squares. For comparison, in the same region
  an all-sky study by \citet{marton2016}, using a statistical method on the WISE and 2MASS
  catalogues, finds 3 of the 9 Class~Is and 3 of the 8 Class~IIs.
See Table~\ref{wise-ysos} for details about the Class~I and Class~II YSOs.  

\subsection{The near-infrared view}
\label{nir-results}

\begin{figure}
  \includegraphics[width=9cm]{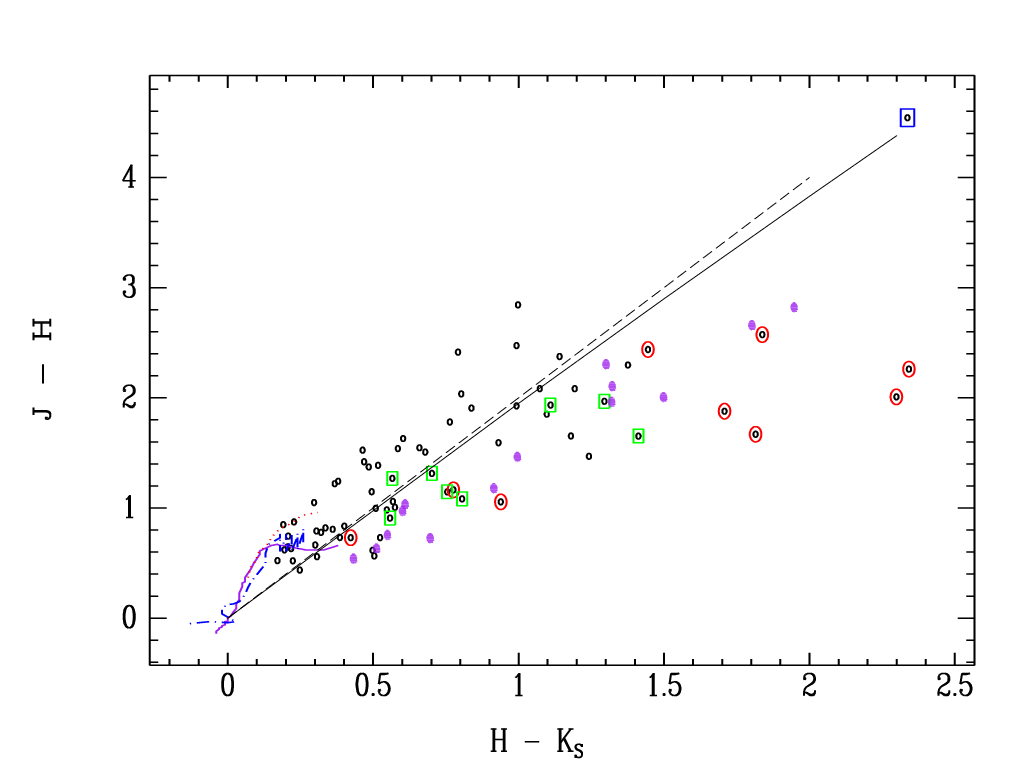}
  \caption{Near-IR $J-H$/$H-K_S$ diagram of 2MASS sources (black circles) where the previously
    classified Class~Is (red) and Class~IIs (green) from WISE colours are marked. Additional
    near-IR photometry from NOTCam WF and the two HR fields that give near-IR excess are added
    (violet dots). For reference, the very bright and red source in Fig.~\ref{notcam-wf}, is
    indicated with a blue square and discussed in Sect.~\ref{sg-results}.
    The loci  of unreddened main-sequence stars, giants and supergiants are indicated with a solid (violet),
    a dotted (red), and a dash-dot (blue) curve. The dashed black line shows the reddening slope of
    an A0 star, and the fulldrawn line shows the reddening curve (see text).    
 }
\label{jhk-diagram}
\end{figure}

Figure~\ref{notcam-wf} shows a color-composed image from broad-band filters J (blue),
H (green), and a narrow-band filter centred on the 2.122 $\mu$m H$_2$ line (red) of the
$4' \times 4'$ region around IRAS~00267+6611 obtained with the NOTCam WF camera. Because of
the very red and bright source in the centre of the field, only very short exposures were
obtained to avoid saturation of the detector. The 2MASS sources in this region are shown in
a $J-H/H-K_S$ diagramme in Fig.~\ref{jhk-diagram}. The loci of unreddened main-sequence
stars, giants and supergiants are indicated with a solid (violet), a dotted (red), and a
dash-dot (blue) curve, respectively, based on their intrinsic near-IR colours tabulated in
the compilation of \citet{tokunaga2000}. We have adopted the near-IR extinction law
$A_{\lambda} \propto \lambda^{-2.07}$ recommended by \citet{wang2019} to be the best average
global extinction law in the wavelength range 1 - 3.3 $\mu$m. This gives a reddening slope
in the $J-H/H-K_S$ diagramme for the 2MASS filters of 2.0, which is depicted with a dashed
line that shows the reddening vector of an A0 star. We note that for highly reddened objects
the extinction law follows a curve in this diagram \citep{kaas1999}, as shown by the
fulldrawn line, where deviations away from the slope is given for the 2MASS colours
by \citet{stead2009}. One particular object in this diagram is sufficiently reddened that
this deviation is not insignificant. This is the source marked by a blue square, the very
brightest object in the K$_S$ band, see description in Sect~\ref{sg-results}. The previously
classified YSOs from the WISE data are high-lighted in this diagramme with red circles
(Class~Is) and green squares (Class~IIs). These sources are listed in Table~\ref{wise-ysos}. 

To further explore the YSO content in the region we use the $J-H/H-K_S$ diagramme to search
for near-IR excess. Bona-fide near-IR excess sources are defined as those that are displaced
to the right and below the reddening vector by more than 2$\sigma$ the errors in the colour
indices. We find 2MASS near-IR excesses in 6 of the previously classified YSOs with WISE,
but only one additional near-IR excess source is found, however, and the reason for this is
the lack of valid entries for the magnitude errors in all JHK$_S$ bands. Some of the sources
located in the near-IR excess region therefore do not qualify as bona-fide near-IR excess
sources. For these we use instead the photometry from the short exposure NOTCam WF-camera
JHK$_S$ images, and in addition, we use the two smaller fields observed with the HR-camera,
their locations are marked in Fig.~\ref{ysos-overlaid} with
dashed yellow and green lines and the images are shown in Fig.~\ref{notcam-hr}.

Due to high extinction there are very few detections in the $J$ band compared to the K$_S$ band,
and this limits the depth to which we can search for near-IR excess sources. Many of the sources
with a large $H-K_S$ colour index, but no detectable $J$-band fluxes, could be fainter YSOs
embedded in the cloud, and some are already found to be Class~Is using WISE, but deeper
$J$-band photometry is needed to reliably distinguish between near-IR excess and reddened
background objects. Point source fluxes are measured to an accuracy of 10\% or better
down to J=19.4~mag, H=18.2~mag and K$_S$=14.7~mag in the WF-camera large field and down
to  J=20.0~mag, H=19.7~mag, and K$_S$=18.7 mag in the HR small fields, but because of uneven
 coverage we estimate the overall completeness to be that of 2MASS at J=16~mag, H=15~mag,
 and K$_S$ = 14.7~mag.
For the NOTCam sample we apply a similar criterion for near-IR excess as above, but we use a
less steep reddening slope of 1.7 for the $J-H/H-K_S$ diagramme, calculated for the NOTCam
filters using the same extinction law \citep{wang2019}.
The NOTCam photometry detects near-IR excesses in several of the previously found Class~I and
Class~II sources and reveals 14 additional sources that we suggest as YSO candidates based on
excess emission at 2 $\mu$m. These are tabulated in Table~\ref{notcam-ysos}, and we have added
them as violet dots to the 2MASS  $J-H/H-K_S$ diagramme in Fig.~\ref{jhk-diagram}, and note that some of them occupy the same locus as Class~I sources.

\begin{figure}
  \includegraphics[width=9cm]{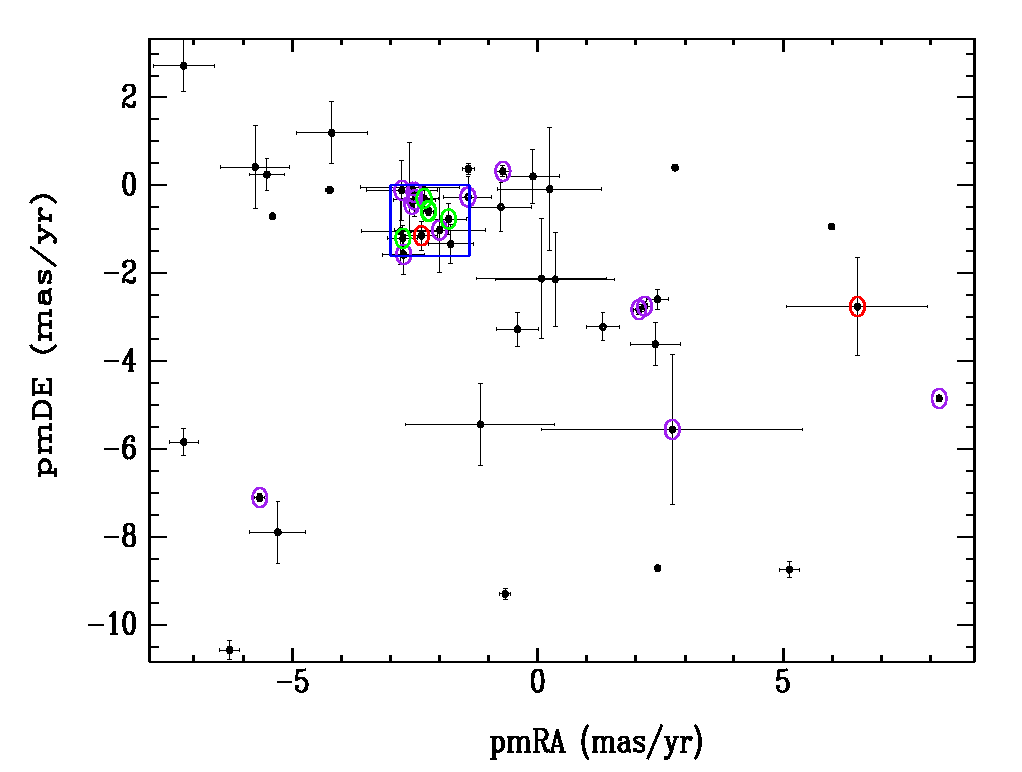}
  \caption{Proper motions of the 57 Gaia DR3 detected stars within the 4$\arcmin \times$
    4$\arcmin$ region of our study (black dots with errorbars). The YSO candidates
    detected are colour coded as follows: red for the single Class~I source, green for 4
    Class~II sources and violet for 13 near-IR excess or variability candidates. Most YSO
    candidates cluster in the blue box.}
  \label{gaia}
\end{figure}

When comparing the photometry obtained with 2MASS and NOTCam some stars stand out as clearly
variable. Addding the magnitude errors in each band from 2MASS and NOTCam in quadrature,
sources that have deviations larger than 3$\sigma$ are flagged as variable. Among 74 common
sources 20 had varied by more than 3$\sigma$ in one of the bands, while 8 sources have
varied in two or more bands. Among the variables are 3 Class~Is, 3 Class~IIs and 5 near-IR
excess sources.

Using variability as a YSO criterion is tempting, as young stars are prone to vary, but
some of the variations we see could be related to the different spatial resolution of the
samples causing blending. In order to weed out false positives we tested the YSO candidates
against Gaia \citep{gaia2016} proper motions. The Gaia DR3 catalogue \citep{gaia2023} lists
57 sources in the 4$\arcmin$ $\times$ 4$\arcmin$ area of our study. Many YSO candidates are
not detected due to being embedded or highly extinguished, but proper motions were found
for 2 of the Class~Is, 5 of the Class~IIs and 13 of the sources selected on the basis of
near-IR excess or variability.
Figure~\ref{gaia} shows the proper motion distribution of all Gaia sources in the region
(black dots with error bars), whereas the YSO candidates are encircled with colour codes.
Most of the YSOs cluster around similar proper motions outlined by the blue box. The box
area has been defined by the individual error bars of one Class I and 5 Class II sources,
and inside the box we find 16 sources. Their median proper motions are $-$2.4 mas~yr$^{-1}$
and $-$0.6 mas~yr$^{-1}$, for pmRA and pmDE, respectively. The scatter in the proper motions
of these sources is 0.41 and 0.48 mas~yr$^{-1}$ in pmRA and pmDE, respectively, which is of
the same order as the individual Gaia measurement errors, the mean error being
0.43~mas~yr$^{-1}$. Thus, we cannot use the proper motion scatter to extract a YSO velocity
dispersion. The intrinsic velocity dispersion is smaller than resolvable by Gaia at this distance,
but we can use 0.4 mas~yr$^{-1}$ as an upper limit (2~km~s$^{-1}$ at $d = 1100$~pc).

Regarding the variability criterion, we find that 6 variable candidates have deviating proper
motions and are discarded, leaving 7 variability selected YSOs of which four are not detected
by Gaia. We add these to the near-IR excess sources, giving a total of 21 YSO candidates from
near-IR photometry, see Table~\ref{notcam-ysos}.
Surprisingly, one Class~I source has a deviating proper motion. We note that this is the
faintest of all the Class~Is at 22 $\mu$m and it is situed outside the Scuba cores
(cf. Fig.~\ref{ysos-overlaid}).
All the YSOs with Gaia measurements have large errors in their parallaxes, except for
two of the Class~IIs we use to estimate the distance (cf. Sect.~\ref{distance}).

A machine learning method to identify YSO candidates from Gaia DR2 and the AllWISE catalogue \citep{marton2019} finds a total of 17 candidates in the field we study. Only 6 of these co-incide with our YSO candidates, all of which have a YSO probability >~0.95 in their catalogue. Three of them are the same Class~IIs as found in \citet{marton2016} and mentioned in Sect.~\ref{mir-results}, and two more Class~IIs are recovered, and in addition one of the YSOs we classify through near-IR variability. The remaining 11 sources do not fullfill our criteria, and only three of these have >~90\% YSO probability in their study.

The final sample of YSOs in the region is shown in Fig.~\ref{ysos-overlaid} overlaid on a
submm map from Scuba/JCMT at 850 $\mu$m, showing the dense cores in the region. The Class~I
protostar positions are marked with large red circles, Class~II sources with green squares,
near-IR excess sources with orange squares and variable sources with cyan diamonds.

\begin{figure}
 \includegraphics[width=9cm]{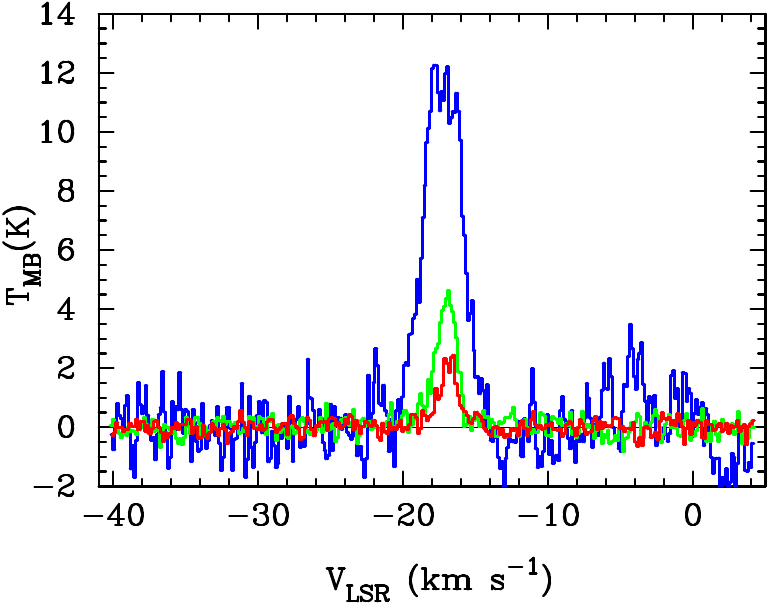}
 \caption{Spectra of the three observed molecular transitions towards the IRAS source
   ((0,0) position of the maps): $^{12}$CO(1$-$0) (blue), $^{13}$CO(1$-$0) (green),
   CS(2$-$1) (red).}
 \label{spectra}
\end{figure}

\subsection{Millimetre emission}
\label{mm-results}

Figure~\ref{spectra} shows the $^{12}$CO(1$-$0), the $^{13}$CO(1$-$0), and the CS(2$-$1)
spectra taken towards IRAS~00267+6511  ((0,0) position of the map). The spectra all
peak at about $-17.0$~km~s$^{-1}$, a value that can be adopted as the velocity of the
gas component associated with the young stellar sources, and are consistent with the
global radial velocity of PGCC~G120.67+02.66 quoted by \citet{zhang2016} in their CO
survey of Planck cold cores in the second Galactic quadrant. They have mapped this
region,  with HPBW of about 52$\arcsec$, using the on-the-fly (OTF) observing mode, with a scan speed of 20$\arcsec$~sec$^{-1}$, and detected strong CO emission. The CO lines were found at $V_{LSR}= -17.56$,
with velocity resolution of 0.17 km s$^{-1}$. HCO+ and HCN emission have also been detected \citep{yuan2016}. 
Our detection of CS emission confirms the presence of high
volume density gas in this region. 

\begin{figure}
\includegraphics[width=8cm]{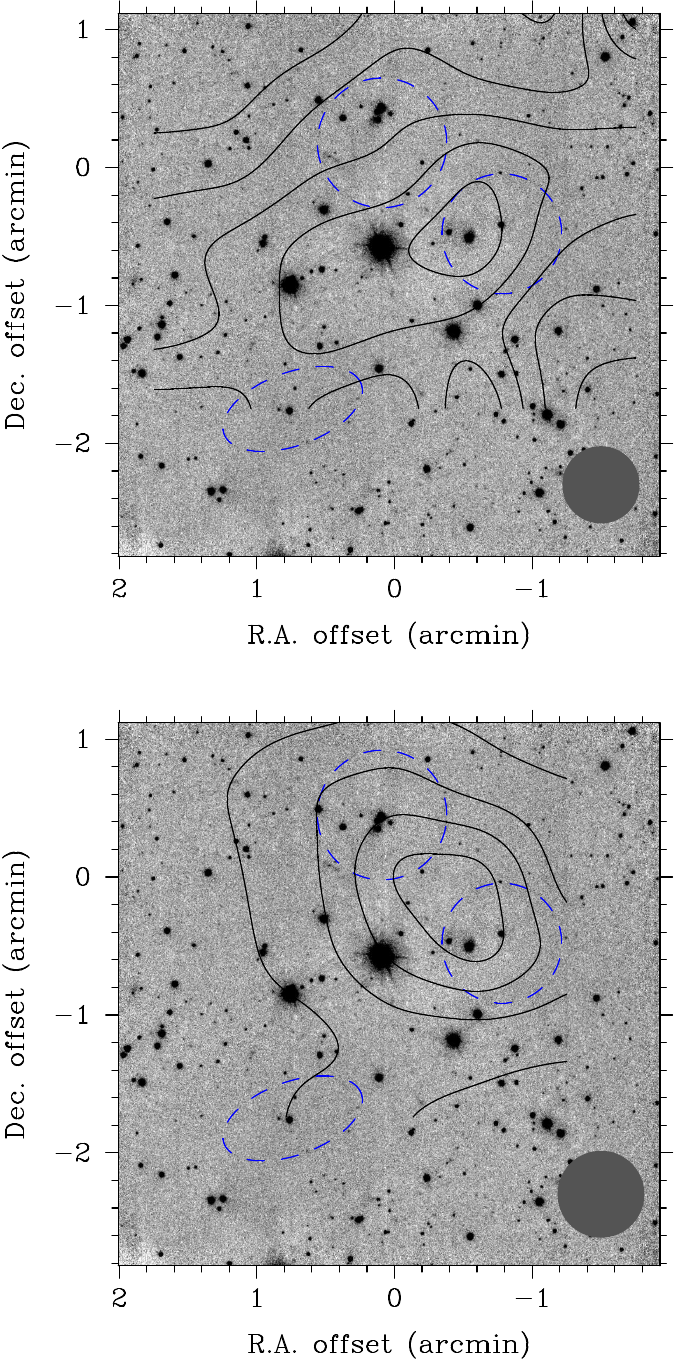}
\caption{Smoothed contour maps of millimetre emission towards IRAS~00267+6511 superimposed
  on the NOTCam H-band image: {\it (top)} $^{13}$CO(1$-$0) integrated intensity: the lowest
  contour is at 6~K~km~s$^{-1}$,
  subsequent contours are in steps of 0.8~K~km~s$^{-1}$ ; {\it (bottom)}: CS(2$-$1) intensity:
  the lowest contour is at 1.0~K, 
  subsequent contours are in steps of 0.4~K. 
  The offsets are relative to the position of the IRAS source. The solid circle represents
  the telescope main beam.  Notice the presence of a high column density peak and plateau
  on the $^{13}$CO map, and a high volume density core on the CS map, both about $0'.7$
  southwest of the IRAS source. The blue dashed contours represent the approximate locations
  and extents of the three brightest SCUBA-2 cores detected towards this region (see later in
  the text). 
} 
\label{13co-cs}
\end{figure}

\begin{table*}
 {\small
 \caption{SCUBA-2 cores in the region around IRAS~00267+6511.}
\label{table:scope}      
\begin{tabular}{l c c c c c c c}    
\hline\hline             
Number & Name &  $\Delta$ R.A. & $\Delta$ Dec & Effective radius & Integrated flux & Core mass & SCOPE name \\
& & (arcmin) & (arcmin)  & (arcmin) & (Jy) & (M$_\odot$) & {\scriptsize  Eden et al. (2019)} \\
\hline                    
\noalign{\vspace{2 pt}}
1 & G120.652+2.681 & $-0.78$ & $-0.48$ & 0.44 & 2.31 (0.23) & 140 (14) & G120.65+2.68 \\ 
2 & G120.668+2.691 & $0.02$ & $0.45$ & 	0.47 & 2.15 (0.20) & 130 (13) &  G120.67+2.69 \\ 
3 & G120.676+2.658 & $0.74$ & $-1.75$ & 0.37 & 1.57 (0.16) & 96 (10) &  G120.68+2.66 \\ 
4 & G120.655+2.645 & $-0.45$ & $-2.62$ & 0.20 & 0.38 (0.04) & 23 (2.0) & $\ldots$ \\ 
5 & G120.689+2.678 & $1.42$ & $-0.48$ & 0.11 & 0.16 (0.01) & 9.7 (1.0) & $\ldots$ \\ 
\hline\hline                              
\end{tabular}
}
\end{table*}

\subsubsection{CO and CS emission}

The $^{12}$CO(1$-$0) maps (both intensity and integrated intensity, not shown) reveal no
remarkable features apart from strong emission with little spatial variation. However,
the values of $^{12}$CO(1$-$0) emission, even though it is usually optically thick,
can be used to derive a first estimate of the optical visual extinction $A_V$ in the
lines-of-sight of the $^{12}$CO(1$-$0) beams. This estimate is based on the empirical
correlation between the $^{12}$CO(1$-$0) integrated intensity and the H$_{2}$ column density
along the line of sight, $N_{i,j}$, at position $(i,j)$:

\begin{equation}\label{nh2}
N_{i,j}(\mathrm{H}_2)=X \int T_{\mathrm{MB}}~dv \qquad (\mathrm{cm^{-2}})\;,
\end{equation}
together with the relation
$N(\mathrm{H}_2)/A_V=9.4 \times 10^{20}$~cm$^{-2}$~mag$^{-1}$ \citep{fre82}.

The constant $X$ has been determined empirically by various works. Adopting the value used
by \citet{may97} for the outer Galaxy, of $X=3.8 \times 10^{20}$~cm$^{-2}$~(K~km~s$^{-1}$)$^{-1}$,
we obtain a first estimate of $A_V=20$~mag.

Figure~\ref{13co-cs} presents the contour maps of $^{13}$CO(1$-$0) integrated intensity and
CS(2$-$1) intensity (black lines) towards IRAS~00267+6511 superimposed on $4' \times 4'$
NOTCam $H$-band images. Both transitions peak about 0.7 arcmin southwest of the IRAS source
revealing the presence of a molecular cloud core with high column densities and high volume
densities in this region. This core is part of the source named PGCC~G120.67+02.66 and mapped
in $^{12}$CO and $^{13}$CO by \citet{zhang2016}, here mapped with higher angular resolution. 

Around the peak of $^{13}$CO emission, there is a relatively slow decrease, with an extended
plateau of emission  (second highest contour level) where column densities remain high. The bright reddened star lies at a
position within this plateau. Using the value of the
$^{13}$CO emission, under the assumptions of optically thin and LTE conditions, we derive a
value of $A_V=6$ mag at the position of this star. This is clearly a lower limit,
because in regions of high extinction the $^{13}$CO(1$-$0) transition is not optically thin
and the calibration methods \citep[e.g.][]{fre82}, that were derived for small values of
$A_V$, do not lead to accurate results when applied in the presence of large values of visual
extinction. 

The CS intensity map displays a single strong peak of emission at approximately the same map
position as the peak on the $^{13}$CO map. However, unlike $^{13}$CO(1$-$0), away from the CS
peak the emission exhibits a strong gradient in all directions around an elongated,
approximately elliptical core with major axis oriented close to the NE$-$SW direction.
Furthermore, a linear fit to the radial CS integrated intensity results in a radial density profile vary according to $r^{-0.6}$, where $r$ is the distance to the peak intensity. Assuming optically thin emission, the column density profile has the same radial dependance. Disregarding any effects of finite cloud radius (or core edge proximity) \citep{yun91}, this implies a volum density radial profile $n(r) \propto r^{-1.6}$, close to $n(r) \propto r^{-3/2}$, the theoretical radial dependance of an infinite isothermal sphere in free-fall collapse  \citep{shu77}.

The YSOs identified on the infrared images lie in regions of the $^{13}$CO and CS maps around the plateau and the peak of maximum emission. Class II sources appear more scattered, whereas most Class I sources are located closer to the peaks of emission (see also Fig.~\ref{ysos-overlaid}), in regions of higher density, supporting their classification as Class I objects (more embedded than those classified as Class II).

\subsubsection{Planck Galactic Cold Clumps and SCUBA-2 cores}

Analysis and photometry of the 850~$\mu$m SCUBA-2 image resulted in the identification of five
sub-mm sources in the region covered by our near-infrared NOTCam images, They are listed in
Table~\ref{table:scope} (positions relative to IRAS~00267+6511, effective sizes, and
850~$\mu$m fluxes). 

The three brightest and more extended sources are also marked by blue dashed lines on
Fig.~\ref{13co-cs}. They coincide with three listed SCOPE cores \citep{eden2019} as indicated in the last column of Table~\ref{table:scope}, and were also detected by the Herschel satellite at 250, 350, and
500~$\mu$m, where the first two brightest are seen as bright sources at all three wavelengths,
and the third source is much fainter. In addition, the large scale Herschel images reveal
that these cores are part of a filamentary structure that include filaments that cross at the
location of these cores (hub filaments). 

The two brightest SCUBA-2 cores have positions on each side of the peak of the CS map along the
major axis of the elliptical CS core (Fig.~\ref{13co-cs}). The brightest core (\#1 in
Table~\ref{table:scope}), also appears to be the densest: its location coincides with the
reddest WISE source, and with the peaks of the $^{13}$CO and CS integrated intensity maps.
Furthermore, it is the region where the CO and CS spectral lines exhibit larger line widths.
Core \#2 coincides with the IRAS position. The two brightest SCUBA cores roughly cover the
NOTCam HR fields N and NW, respectively, seen in Fig~\ref{ysos-overlaid}. 

Because dust continuum emission at sub-mm wavelengths is optically thin, we can estimate the
masses of the SCUBA-2 cores. Under the assumption of constant dust temperature, the dust mass
can be obtained using: 
$$  M_{\rm dust}=\frac{S_{\lambda}}{\kappa_{\lambda} B_{\lambda} (T_d)}  D^2   $$
\noindent
where $S_\lambda$ is the 850 $\mu$m flux, $T_d$ is the dust
temperature ($T_d = 12.6$~K, \citet{planck2016}), and $\kappa$ is the mass absorption coefficient (or emissivity) of the dust. We adopted a distance of 1.1~kpc (See
Sect.~\ref{distance}), a dust-to-gas ratio of 1\%, and a spectral index $\beta=2$ for the emissivity
law $\kappa_{\lambda} \propto \lambda^{-\beta}$, with $\kappa_{\rm 1 mm}=0.3$~cm$^2$ g$^{-1}$.
The derived masses of the SCUBA-2 cores are given in the last column of Table~\ref{table:scope}.

\bigskip
\subsubsection{CO and CS parameters of the SCUBA-2 cores}

Table~\ref{table:scope-cocs} lists the CO and CS line parameters of each of the SCUBA-2 cores that lies within the CO or CS maps. The values listed are those of the CO or CS beams closest to the centres of each core.

\begin{table}[h]
 {\small
 \caption{CO and CS parameters of the SCUBA-2 cores.}
\label{table:scope-cocs}      
\begin{tabular}{c l c c c}    
\hline\hline             
 & & \multicolumn{3}{c}{Core number} \\
  \cline{3-5}

 & & 1 & 2 & 5 \\

 \hline                    
\noalign{\vspace{2 pt}}
CO & $T_R$ (K) & 11.4 & 11.4  & $\ldots$ \\ 
 & $\int{T_R dv}$ (K km s$^{-1}$) & 43.0 & 35.3 & $\ldots$  \\ 
 & $v_{\rm LSR}$ (km s$^{-1}$) &  $-$18.11 & $-$17.12  & $\ldots$  \\  
 & $\Delta v$ (km s$^{-1}$) & 3.53 & 2.92 & $\ldots$ \\
\noalign{\vspace{2 pt}}
$^{13}$CO & $T_R$ (K) & 3.6 & 4.6  & 4.5 \\ 
 & $\int{T_R dv}$ (K km s$^{-1}$) & 7.7 & 7.3 & 7.2  \\ 
 & $v_{\rm LSR}$ (km s$^{-1}$) &  $-$18.30 & $-$16.89  & $-$17.23  \\  
 & $\Delta v$ (km s$^{-1}$) & 1.99 & 1.49 & 1.51 \\
 \noalign{\vspace{2 pt}}
CS & $T_R$ (K) & 1.57 & 1.58  & $\ldots$ \\ 
 & $\int{T_R dv}$ (K km s$^{-1}$) & 4.2 & 1.96 & $\ldots$  \\ 
 & $v_{\rm LSR}$ (km s$^{-1}$) &  $-$18.32 & $-$16.55  & $\ldots$  \\  
 & $\Delta v$ (km s$^{-1}$) & 2.52 & 1.17 & $\ldots$ \\
\hline\hline                              
\end{tabular}
}
\end{table}

\bigskip
\section{Discussion}
\label{discussion}

\subsection{Distance}
\label{distance}

From the spectral lines, we derive the kinematic heliocentric distance, using the peak
radial velocity. This can be done using different methods namely, e.g.\ applying the
circular rotation model by \citet{brand93}, or using the \citet{reid2014} rotation curve
with the \citet{reid2019} updated Solar motion parameters, possibly including Monte
Carlo methods \citep[e.g.][]{wenger2018}, among others.
Using the conventional solar motion parameters, and with $V_{\mathrm{LSR}}=-17$~km$^{-1}$,
a distance of 1.5~kpc is obtained, not far from the value of 1.75~kpc derived by
\citet{zhang2016} using the \citet{sofue2011} method. \citet{zhang2018} using a Bayesian
calculation derive a distance of $0.90 \pm 0.29$ kpc to G120.67+02.66.
However, if we adopt the \citet{reid2014} rotation curve together with the revised values
of the Galactic constants $R_0$ and $\Theta_0 $ of \citet{reid2019}, we derive distances
between 0.95 and 1.25 kpc. Interestingly, \citet{guo2020} use data from extinction
catalogues and parallaxes from Gaia DR2, together with modelling, and
derive distances to Planck Galactic cold cores. They obtained a distance of $1063~\pm~104$~pc
to PLCKECC G120.67+02.66. Furthermore, we find that among the 8 Class~II sources found
in Sect.~\ref{mir-results}, as many as 5 have measured parallaxes in the GAIA DR3 catalogue
\citep{gaia2023}, and two of these have relatively small errors. These two optically visible
Class~IIs are located relatively close to the center of the region, and we use their
parallaxes at 0.867 $\pm$ 0.025 mas and 0.948 $\pm$ 0.097 mas to estimate a distance of
1.1 $\pm$ 0.05 kpc. We conclude that the currently best estimate for the distance to this
star formation region and molecular cloud core is 1.1~kpc.

\begin{figure*}
  \includegraphics[width=9cm]{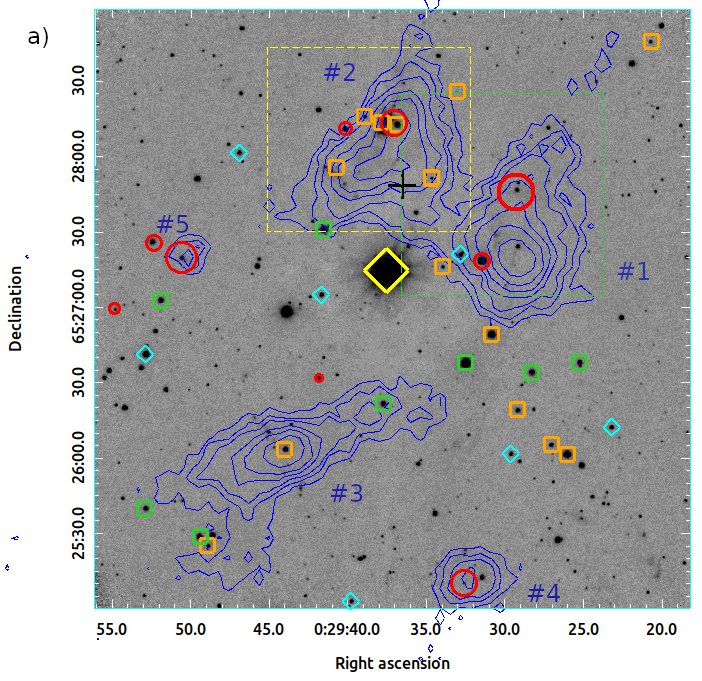}
 \includegraphics[width=9cm]{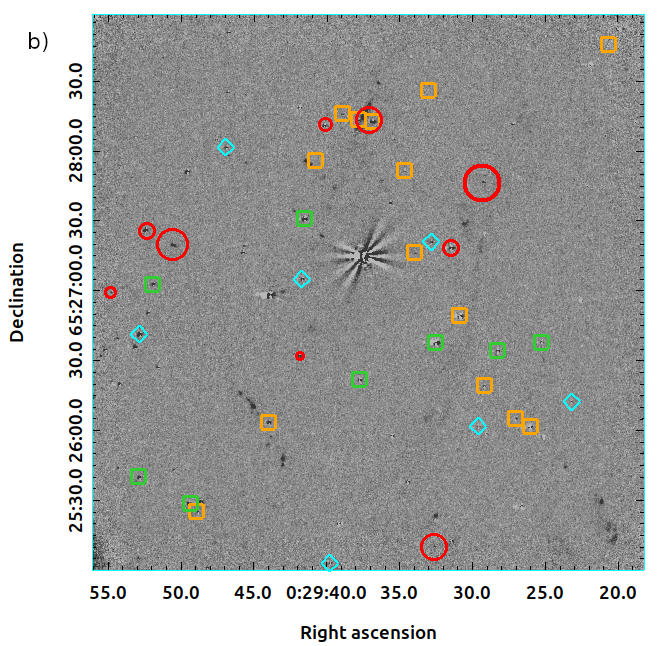}
 \caption{ a) The YSO sample is overlaid on the H$_2$ (2.122$\mu$m) narrow-band image with
   the position of  IRAS~00267+6511 marked (black plus sign). The SCUBA-2 cores are numbered as
   in Table~\ref{table:scope}, and blue contours show 850 $\mu$m continuum emission in
   steps of 0.2, 0.3, 0.4, 0.6, 0.8, 1.1 and 1.6 mJy per square arcsec. Positions of YSOs are marked
   for Class~I sources (red circles, where size suggests relative brightness at 22~$\mu$m),
   Class~II sources (green squares), near-IR excess sources (orange squares) and variable sources
   (cyan diamonds). The putative background supergiant (yellow diamond).
   The dashed boxes outline the small regions shown in Fig.~\ref{notcam-hr} at higher resolution.
    b) Continuum subtracted H$_2$ (2.122$\mu$m) image with YSO positions marked as in panel
   a). Some artefacts are due to stellar PSFs not well subtracted. (Note also the diffraction spikes from
   the telescope spider at the  bright star as the telescope field rotator adjusts.)
    }
  \label{ysos-overlaid}
\end{figure*}

\subsection{The embedded YSO population}
\label{yso-pop}

The results from exploring the stellar population with mid-IR WISE and near-IR 2MASS and
NOTCam photometry as explained in Sect.~\ref{mir-results} and Sect.~\ref{nir-results}
demonstrate a rich collection of young stars with 9 Class~I sources, 8 Class~II sources,
and 21 YSO candidates from the near-IR study. The total sample of 38 YSOs is shown in
Fig.~\ref{ysos-overlaid} overlaid on the 4$\arcmin \times$ 4$\arcmin$ narrow-band
2.122~$\mu$m H$_2$ image which for a distance of 1.1~kpc corresponds to an area of
1.3 pc $\times$ 1.3 pc, giving an average stellar surface density of 24~YSOs~pc$^{-2}$.
This would meet the criterion of a stellar cluster as defined in \citet{lada2003}, who define it as a group of 35 or more physically related stars whose stellar
  mass density exceeds 1.0~M$_{\odot}$~pc$^{-3}$.

The contours in the left panel of Fig.~\ref{ysos-overlaid} show the high spatial
resolution 850~$\mu$m
mapping from Scuba/JCMT that outlines the dense cores listed in Table~\ref{table:scope}.
We note that all cores are populated with YSOs, and that most Class~Is are embedded in a
core. Even the two smallest cores have a Class~I source inside the density contours. While
the spatial distribution of the YSOs in general is relatively scattered, the Class~I sources
are essentially located in the cores and the remaining YSOs partly in the submm cores and
partly outside. 

The number ratio Class~I/Class~II is a measure of activity and youth since the average
lifetime of the Class~I stage is estimated to 0.4~Myr \citep{evans2009} and the
duration of the Class~II phase is typically taken to be 2-3~Myr \citep[see also][]{dunham2015}.
This means that a number ratio around unity suggests the region is extremely young with
on-going formation of protostars. Such high ratios are typical in environments of tight stellar
groups according to the {\em Spitzer} based study of the solar neighbourhood \citep{evans2009}.
\citet{gutermuth2009} finds a median Class~I/Class~II ratio of 0.27 in 36 star-forming cores
within 1 kpc of the Sun. The 21 YSO candidates from the near-IR study cannot be classified
as bona-fide Class~IIs, but most are likely pre-main sequence disk sources, although
Fig.~\ref{jhk-diagram} shows that some have very red $H-K_S$ indices and occupy the region of
Class~Is, and these are too faint and/or not resolved in the WISE sample. If we would consider to add these 21 sources to the Class~II category, we would obtain a lower limit
number fraction of 9/29 over the whole field, which is still high.

Another indication of youth is the location of the YSOs with respect to their birthplaces, presumably
in the cores. Stellar velocity dispersions obtained from radial velocity measurements in clusters with massive stars give values as high as 3.5~km~s$^{-1}$ for NGC~2264 \citep{fhuresz2006} and  3.3~km~s$^{-1}$ for Berkeley~59 \citep{gahm2022}, while based on Gaia DR2 \citet{kuhn2019} find lower dispersions for these same clusters and typical values of 1-3~km~s$^{-1}$ for clusters in general. In Sect.~\ref{nir-results} Gaia proper motions give us an upper limit of 2~km~s$^{-1}$.  For embedded clusters, near-IR radial velocity studies find velocity dispersions of 0.9~km~s$^{-1}$ for NGC~1333 \citep{foster2015}, a cluster which is similar to this region in many aspects; it forms low-mass stars, is embedded and young enough to contain both dense cores and protostars. With such velocities the YSOs would disperse by 0.9~pc in the time span of 1~Myr, i.e. they would escape the dense cores in less than 0.5~Myr.
  These considerations suggest that the age of this star formation region is likely below 1 Myr.

Given the embeddedness and the very early stage of star formation, we have not
attempted to determine the masses of the YSOs.

The SCUBA-2 core \#2 in Table~\ref{table:scope} (G120.668+2.691, the northern-most in
Fig~\ref{ysos-overlaid}) contains 8 identified YSOs, including two bona-fide Class~I protostars,
and is the core richest in detected YSOs. For a 1.1~kpc distance the effective core radius is
0.15~pc, giving a stellar surface density of 113~YSOs~pc$^{-2}$ ($\pm 10\%$), which is
similar to that of the central cluster (A) in the Serpens Cloud Core measured within an edge
of A$_V$ = 20 mag \citep[See][Table 6, adjusted to d=415 pc]{harvey2007}.
Assuming symmetry in the radial direction, the volume density in this core is 570~YSOs~pc$^{-3}$.
This fits well with the definition of a tight stellar group (~$>\,25~M_{\odot}$~pc$^{-3}$)
by \citet{evans2009} even if we would assign a low average stellar mass.
If we assume that the mean YSO mass is 1~M$_{\odot}$ and use the estimated core mass in
Table~\ref{table:scope}, the present star formation efficiency (SFE) in this core 
reaches $\sim$~6\%.

\begin{figure*}
  \includegraphics[width=5.8cm]{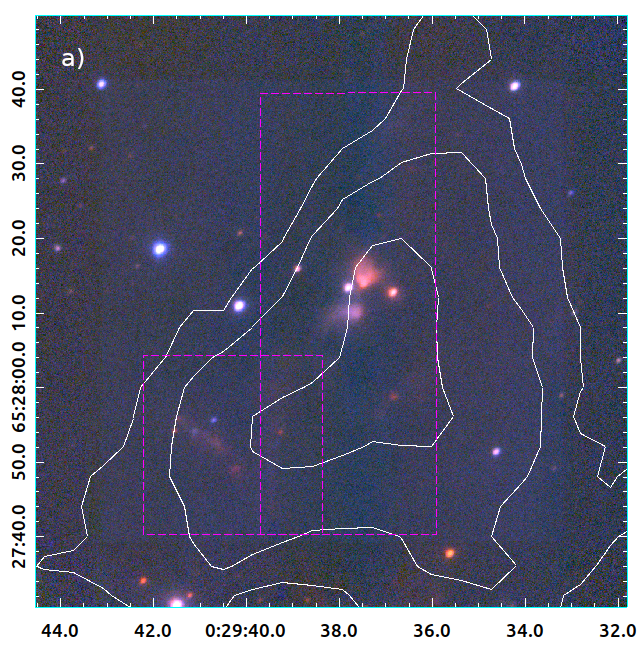}
  \hspace{1mm}
     \includegraphics[totalheight=5.8cm]{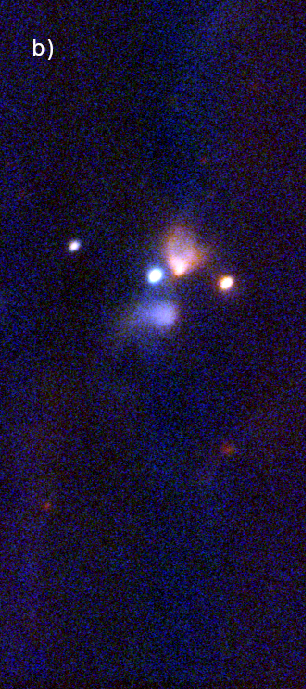}
       \hspace{0.5mm}
   \includegraphics[height=5.8cm]{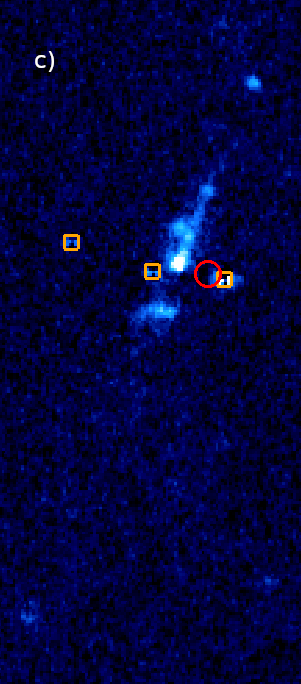}
   \hspace{8.5mm}
   \includegraphics[width=5.8cm]{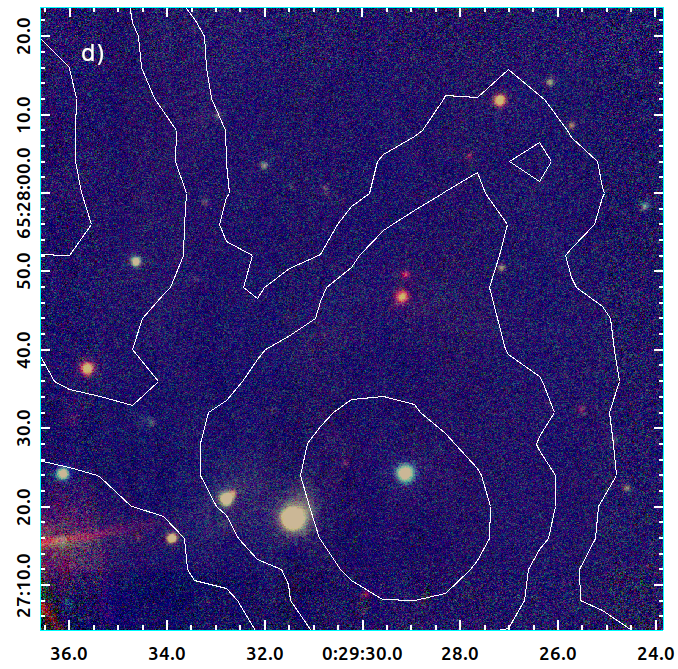}
  \caption{a) and d) shows NOTCam J (blue), H (green) and K$_S$ (red) RGB images of the two
    smaller fields {\em N} and {\em NW}, both 80$\arcsec \times$ 80$\arcsec$
    (0.43$\times$ 0.43 pc) wide. The 850 $\mu$m emission defining the submm cores is
    shown (white contours) in steps of 0.2, 0.4, and 1.0 mJy per square arcsec. The rectangle
    marked in a) is the hour-glass shaped nebulosity shown in JHKs colours in b) and with a
    collimated  jet emerging in the H$_2$ image in c). The tabulated
    position of the WISE Class I source (red circle) and three near-IR excess sources (orange squares)
    are marked in c).}
   \label{notcam-hr}
\end{figure*}

\subsection{Extended emission in the dense cores}
\label{extended-sources}
  Figure~\ref{ysos-overlaid}~b) shows the continuum-subtracted
  emission in the H$_2$ v=1-0 S(1) line at 2.122~$\mu$m. This line is a good tracer of
  shock-excited emission in low-velocity shocks, and frequently used to study embedded
  Herbig Haro flows and protostellar jets from ground-based observations
  \citep[e.g.][and references therein]{reipurth2001}. The figure indicates several flows in
  the area of our study, at least
  three major flows, as well as a number of more isolated emission features.
  The driving sources of such embedded flows are typically found to be protostars. Velocity
  information of the jet structures is required to properly designate the origin of the jets, but
  we can by morphology and proximity relate Class~I sources or near-IR excess YSOs to most
  of the structures, as discussed in the following.

\subsubsection{SCUBA-2 core G120.668+2.691 (\#2)}

SCUBA-2 core G120.668+2.691 (\#2 in Table~\ref{table:scope}) is imaged at high-resolution
in the near-IR in Fig.~\ref{notcam-hr}~a). This core
contains a nebulosity with emission extending out from a dark center forming an hour-glass
shaped nebula, which is seen in our images to extend over at least 7$\arcsec$ across the sky,
corresponding to $\sim$ 8000 au. The structure is seen in all three JHK$_S$ filters, is
strongest in the K$_S$ band, and reddest in the northern part.
Such a morphology can be interpreted as scattered light from cavities formed as a bipolar
jet/wind/outflow from a protostar has begun to clear the dust cocoon in the polar regions.
Although at a lower spatial resolution, this object looks remarkably similar to the nearby
Class~0 protostar L\,1527~IRS as imaged by NIRCam at the James Webb Space Telescope
\citep[See Fig.1 in][]{tobin2024}. The inner region of the northern cone has a sharp peak
which is, however, not a point source as it is slightly more extended (fwhm = 0.7$\arcsec$)
than the stellar PSF (fwhm = 0.4$\arcsec$).  As shown in Fig.~\ref{notcam-hr}~b) and c)
this intensity peak in the K$_S$ image is mainly due to a bright knot of H$_2$ emission, and
 a clear distinction is seen between the cone morphology in broad band JHK$_S$ images and
a narrow bipolar jet, with knots extending over a large distance, traced by the H$_2$ line emission.
In panel c) the position of 
the Class~I source J002937.06+652813.1 from the WISE catalog is marked. It is offset from the
near-IR source next to it; the 2MASS cross-correlation measure gives an offset of 1.45$\arcsec$
together with an extended source flag of 5, which means it falls within 5$\arcsec$ of an
extended 2MASS source. Taking into account the WISE spatial resolution of 6$\arcsec$ and
the high YSO density, the Class~I source is likely the driving source giving rise to the nebula
and the jet, having its catalog coordinates confused with the 2MASS extended source
and the near-IR excess YSO to the west, which, however, could also be a Class~I,
  judging from its location in the $J-H/H-K$ diagramme.
The jet is clearly bipolar with a northern extent of at least 20$\arcsec$, limited by
coverage, and a southern arm of $\sim$ 50$\arcsec$, judging from the furthest aligned knot,
corresponding to about 0.27~pc. If we assume a typical average velocity of protostellar
jet knots to be 50-100~km~s$^{-1}$, as found in a multi-epoch proper-motion study in
Serpens \citep{djup2016}, the dynamical age of this jet would be a few 10$^3$ yr.

\begin{figure}
  \includegraphics[width=4cm]{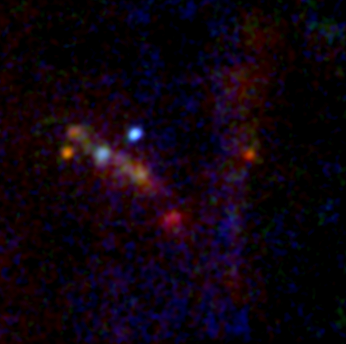}
  \hspace{1mm}
  \includegraphics[width=4cm]{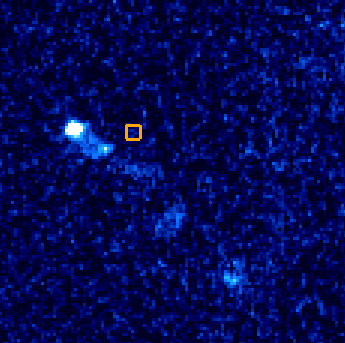}
  \caption{Enhancement of the jet structure in the dashed box from Fig.~\ref{notcam-hr}~a), and
    comparison between JHK$_S$ RGB coded image (left) and continuum subtracted H$_2$ image
    (right). Near-IR excess YSO marked with orange square.}
   \label{jet2}
\end{figure}

Further to the South in field {\em N} several features of faint and red extended emission
are seen. One particular structure is an NE/SW elongated nebulosity that looks like parts
of a jet, but what could be its originating source is not obvious, see Fig.\ref{jet2}.
We found one YSO nearby from near-IR excess, and two additional red point sources
could be deeply embedded YSOs, but failed to meet the classification criteria.

\subsubsection{SCUBA-2 core G120.652+2.681 (\#1)}

Most of SCUBA-2 core G120.652+2.681 (\#1 in Table~\ref{table:scope}) is covered by field
{\em NW} and shown in Fig.~\ref{notcam-hr}~d). This is a very dense region and we
detect  only 4~YSOs (cf. Fig.~\ref{ysos-overlaid}), but it includes the brightest 22~$\mu$m
source, a Class~I protostar (WISE J002929.28+652746.0) that is deeply embedded and has
an equally red but fainter neighbour. We note a curved tail of faint H$_2$ emission extending
from this Class~I  position towards the south.
Another Class~I in this core is located in a group of YSOS in the lower
part of the image in  Fig.~\ref{notcam-hr}~d) and has some extended emission. 
The star located near the peak of the dense submm core is a foreground object. 

\subsubsection{SCUBA-2 core G120.676+2.658 (\#3)}

The southern of the three largest SCUBA-2 cores (\#3, G120.676+2.658), is more elongated and
Fig.~\ref{ysos-overlaid}~a) shows that five YSOs are related to it, none of them
bona-fide Class~I sources. From the continuum-subtracted narrowband 2.122~$\mu$m
H$_2$~line emission image in  Fig.~\ref{ysos-overlaid}~b) we see a jet-like structure of
several knots extending out from the near-IR excess source 2MASS 00294396+6526034, see
Table~\ref{notcam-ysos}, likely the driving source of this jet.

\bigskip

\subsubsection{The two smallest SCUBA-2 cores (\#4 and \#5) }

 Figure \ref{ysos-overlaid}~a) shows that the two smaller submm cores also host
  Class~I protostars. Towards core \#4 in Table~\ref{table:scope} the Class~I source
  WISE J002932.58+652510.0 is located, and a knot of H$_2$ emission is found $\sim$ 13$\arcsec$
  to the North of it. Similarly, the Class~I star WISE J002941.83+652631.9, which we flag in
  Table~\ref{wise-ysos} as a source with a deviating proper motion, could be related to an H$_2$
  emission knot found to the north of it, at a projected distance of approximately 0.07 pc. 
  Finally, the Class~I source (WISE J002950.58+652719.6) is seen towards the submm core \#5,
  and seems to possess a small jet at the very vicinity of the star. 

  \medskip
  
  Some structures are more difficult to interpret, such as the relatively bright features in the SW corner
  of the image. However, an extensive investigation of these jets is outside the scope of the paper.
  We note that the mere presence of such a rich collection of shocked molecular gas features, most
  likely caused by several protostellar jets and flows, supports our conclusion that the region is at an
  early evolutionary star formation phase, comparable to nearby star formation regions such as
  NGC~1333 \citep{davis2008}. Traced by the mass loss through collimated jets, all the
  submm cores in this Planck cold clump host YSOs that have gone through recent major mass
  accretion activity on a similar time-scale.

\subsection{A red supergiant star?}
\label{sg-results}

The brightest source in the infrared images, both in Fig.~\ref{notcam-wf} and in the WISE
atlas is the source 2MASS~00293749+6527146 aka WISE~J002937.50+652714.7, a hindrance for
deep near-IR imaging due to its strong $K$ band brightness. It is located close to the
centre of the region, has an extremely red colour in the near-IR, while nothing
is seen in the optical except for a marginal detection in the PanSTARRS DR1 y-band image,
and it is not in the Gaia DR3 catalogue.
In the near-IR colour-colour diagramme in Fig.~\ref{jhk-diagram} the location of
the source (open blue square) is consistent with it being a highly reddened supergiant or
giant. In the WISE CC diagram in Fig.~\ref{wise-cc} the source has excess emission in the
$[3.4]-[4.6]$ index but not in the $[4.6]-[12]$ index. It also has a clear excess in a
$K_S - [3.4]/[3.4]-[4.6]$ diagram, where it occupies a locus similar to that of the
Class~II sources, but according to the criteria of \citet{koenig2014} the source would be
rejected as a YSO because it is too bright at 3.4 $\mu$m, as it by their definitions would
enter the locus typically occupied by AGB stars. The source was also observed by the
Midcourse Space Experiment (MSX), and the A band (8.28 $\mu$m) flux in the MSX6C point 
source catalogue \citep{egan2003} is 0.59$\pm$0.02 Jy which translates to $A = 4.99$~mag
using 58.49~Jy for zeroth magnitude \citep{egan1999}. 

The NOTCam photometry obtained in 2023 shows that the source has faded by about 0.2 mag in
all bands since the 2MASS measurement obtained in 1999, see Table~\ref{sg-tab}. The MSX A
band variability flag indicates that the source has varied over the course of the MSX
mission (1996-1997) by more than 3$\sigma$. The {\it WISE} variability flag is 2 for the
[3.4] and [4.6] bands, which means a small probability of being variable in the three
epochs over the WISE mission.

\begin{table}
\small{  
  \caption{IR photometry of the red supergiant candidate.}
  \label{sg-tab}
  \begin{tabular}{lccc}
  \hline
  \hline
     &    $J$     &     $H$     &    $K_S$  \\
             & (mag)    &   (mag)   &   (mag)      \\
  \hline
  2MASS  & 13.77$\pm$0.03 & 9.23 $\pm$ 0.03 & 6.89 $\pm$ 0.02  \\
  NOTCam & 13.97$\pm$0.02 & 9.54 $\pm$ 0.03 & 7.09 $\pm$ 0.05  \\
  \end{tabular}
  \begin{tabular}{lcccc}
  \hline  
        &   [3.4]     &    [4.6]     &   [12] &   [22] \\
              & (mag)    &   (mag)   &   (mag) & (mag)     \\
  \hline
  WISE        & 6.06$\pm$0.10 & 5.19 $\pm$ 0.07 & 5.13 $\pm$ 0.02 & 4.61 $\pm$ 0.03 \\
  \hline
  \end{tabular}
}  
\end{table}

\begin{figure}
\includegraphics[width=10cm]{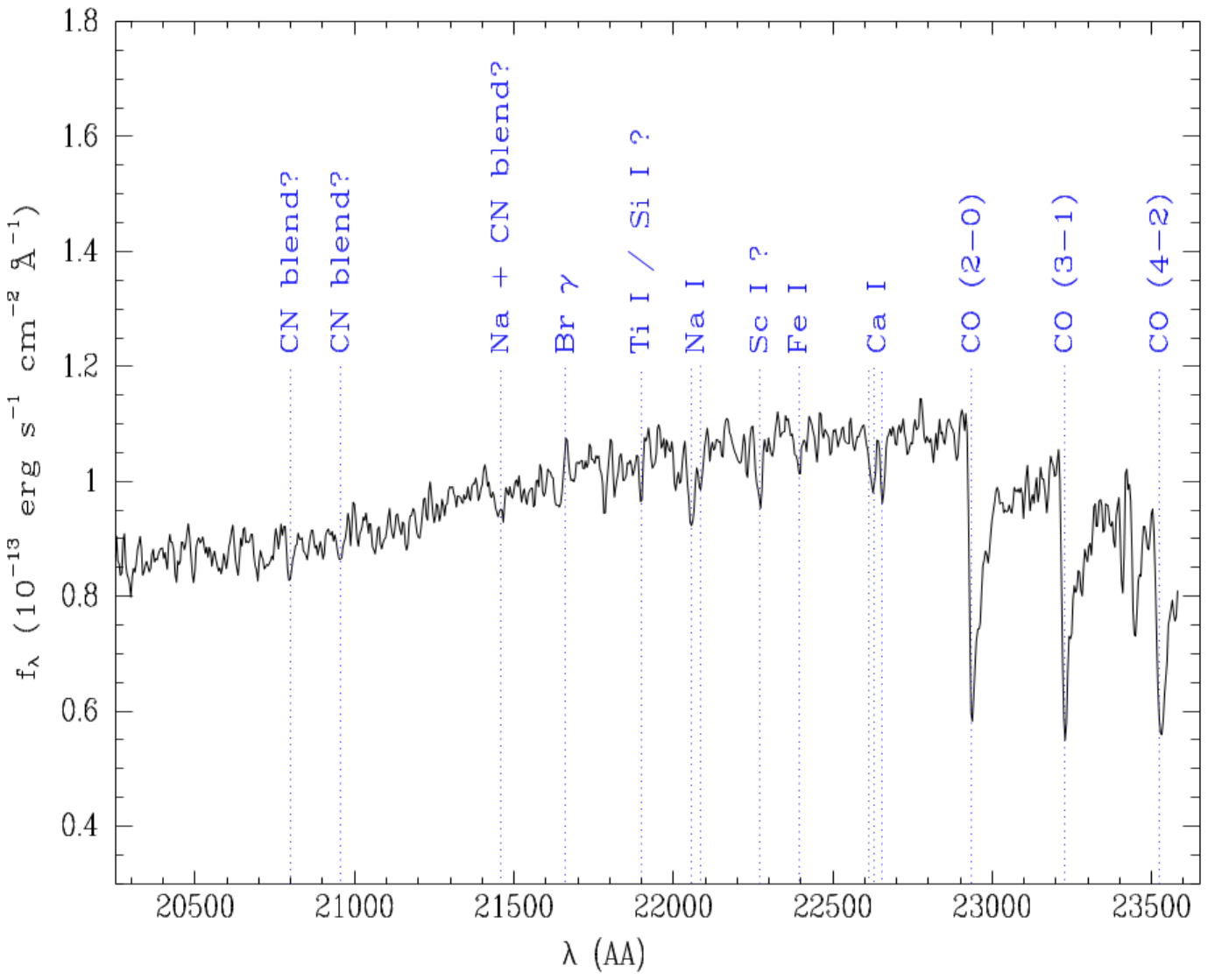}
\caption{NOTCam K-band spectrum of the red luminous star showing the telltale signature of
  a late type star; very strong CO bandhead absorptions, as well as the Ca~I triplet at
  2.263 $\mu$m and the Na~I doublet at 2.207 $\mu$m in absorption. }
\label{kspec}
\end{figure}

A K-band spectrum of the source is shown in Fig.~\ref{kspec}. Its prominent and deep CO
bandhead absorptions at 2.2935 $\mu$m ($2-0$), 2.3227 $\mu$m ($3-1$), and 2.3535 $\mu$m
($4-2$) together with the absorptions of the Ca~I triplet around 2.263 $\mu$m and the
Na~I doublet around 2.207 $\mu$m, and no appreciable Br$\gamma$ at 2.166 $\mu$m, concur
to indicate a late spectral type. We measure the equivalent widths (EWs) as described in
\citet{comeron2004} and find the EWs of the Na~I doublet (2.2062, 2.2090 $\mu$m) to be
5.1 \AA \ and the Ca~I triplet (2.2614, 2.2631, 2.2657 $\mu$m) to be 4.7 \AA. The
reference continuum was measured on both sides of the features and the errors are
estimated to be $<$ 1 \AA . We have measured the EW of the 2.29~$\mu$m CO ($2-0$)
absorption feature from 2.2920 to 2.3020 $\mu$m, i.e. over wavelength range of 0.010
$\mu$m, using reference continuum levels on the blue side of the feature extrapolated to
the red side, and we find EW$_{\rm CO(2-0)}$ =  $25.5 \pm 1$~\AA .
These measured EWs are very similar to those measured for red supergiants
\citep[e.g.][]{comeron2004,comeron2016}. Using the relation between spectral type and the EW of
CO~($2-0$) established in \citet{davies2007}, which is based on the spectral type versus
temperature calibration of \citet{levesque2005}, a tentative IR spectral type of M2~I can
be assigned to the target. We caveat that a later M-type giant could also reach similarly
deep CO absorptions, although in that case the water steam bands are often prominent, and
we do not see any notable depression shortwards of 2.1 $\mu$m in this target. Comparison
was made with the near-IR spectral atlases by \citet{wallace1997} and \citet{lancon2000}.
With help of the online archive of spectra of luminous cool stars in numerical form
provided by \citet{lancon2000}, we could measure the EW of the 2.29~$\mu$m CO feature in
the same manner as for our target for two M2 supergiants and one M5.5 giant and find that
our target has an EW in between the two types. 

The star's auspicious position at the very centre of the star formation region raises
the question of whether it could be an embedded outburst object. We can probably exclude
that it is an FU~Orionis outburst, although they present similarly deep CO absorption bands
and are often bluer when in outburst \citep{szegedi2020}, because its luminosity would be an
order of magnitude larger than what is typical for FUORs, and furthermore, the Na~I and Ca~I
absorptions in the K-band spectrum are too strong for such an interpretation
\citep{connelley2018}.

According to \citet{elias85} the intrinsic
$H-K_S$ colour for the M2~I spectral type is (H-K$_S$)$_0$ = 0.22 mag, while that of an
M5.5~III spectral type is 0.29 mag \citep{bessell1988}, see compilation by
\citet{tokunaga2000}). This gives quite a similar a colour excess in both cases,
$E(H - K_S)$ = ($H -K_S)_{\rm obs} - (H-K_S$)$_0$, of 2.12~mag and 2.05~mag for M2~I and M5~III,
respectively. From the \citet{wang2019} near-IR extinction law the relation
A$_K$ = 1.33~$\times~E(H-K_S)$ gives A$_K$ = 2.82~mag and 2.72~mag for M2~I and M5~III,
respectively. This translates to around 36 magnitudes of visual extinction using 
A$_K$ = 0.078 A$_V$, i.e. the source is extremely reddened even though its projected
location is just offset from the densest submm cores, see Fig.~\ref{ysos-overlaid}.
The expected absolute magnitude of an M5 giant is M$_V$ = --~0.3~mag and that of an M2
supergiant is M$_V$ = --~5.6 mag \citep{cox2000}. The respective intrinsic colours
from the above references are $V-K$ = 5.96 mag and gives M$_K$ = --~6.3 mag (M5 giant)
and $V-K$ = 4.1~mag giving M$_K$ =  --~9.7~mag (M2 supergiant). We can make a rough
estimate of the distances required to reach such absolute K-band magnitudes via the
relation M$_K$~=~m$_K - A_K - 5 \log d - 5$, and we find 1.2~kpc and 5.7~kpc for the
M5 giant and the M2 supergiant, respectively. Thus, we cannot use the distance
argument to exclude the possibility that the source is a late M-type giant. The measured
extinction requires the source to be behind the star formation region at 1.1~kpc. If
it is a giant, however, the uncertainties in the above reasoning would put it at
a distance compatible with the cloud, but owing to the advanced age of a giant, it would
be unlikely that it is related to the cloud.

\subsection{Star formation scenario}
\label{sf-hist}

    The G120.69+2.66 cold clump, which appears to be the central hub of a larger
     Herschel filament, contains several dense submm cores, all of which host protostars
     and/or signs of on-going low-mass star formation. 
     On the other hand, it has been suggested \citep{tang2018,yi2018,zhang2018} that Planck cold clumps have low levels of star formation activity, while  \citet{wu2012} and \citet{liu2013} indicated that they are more quiescent and less evolved than the typical star-forming region.
The values of the radii, masses, and mass surface densities of these cores in G120.69+2.66 fit within the range of the corresponding derived values for compact sources in the recent study of Galactic Plane Planck cold clumps \citep{eden2024}, and thus indicate that these cores are typical members of the family of compact sources identified in Galactic Plane Planck cold clumps. However, the submm cores in G120.69+2.66 have radii (0.04 to 0.15 pc) that place them at the shorter end of the distribution of sizes of the \citet{eden2024} compact sources, and have mass surface densities that are at the higher end of the distribution of mass surface densities. This appears to indicate that G120.69+2.66 belongs to the group of more active Planck cold clumps that are currently sites with on-going star formation. 
 
     In the previous sections we
     argue that the star formation process must be in an early stage, and in fact \citet{juvela2018} found that any correlation of Planck cold clumps with young stellar objects tends to be the youngest protostellar stages. Here we compare
     with well-studied, nearby low-mass embedded clusters that exhibit similar sub-structure
     and dense cores, co-existing with protostars and more evolved YSOs. Several such regions
     were studied by the Spitzer c2d survey
     \citep[e.g.][and references therein]{evans2009}. 
     An inspection of the youngest
     embedded clusters, NGC~1333 at a distance of 296~pc \citep{kuhn2019} and Serpens
     Main at 415~pc \citep{dzib2010}, show many similarities with G120.69+2.66.
     \citet{lada2003} described these as the least evolved embedded clusters known and likely
     no more than 1~Myr of age.

     The Class~I sources in G120.69+2.66 are practically all seen projected on to dense submm
     cores. This was shown for Serpens Main \citep[e.g.][Fig.~11]{kaas2004} and is particularly
     well demonstrated in NGC~1333 \citep[][Fig.~10]{gutermuth2008}, where the  Class~I
     sources perfectly trace the dense submm cores, giving a picture of the pristine early phase
     of the formation of a cluster.
     Counting those YSOs for which we have bona-fide mid-IR classification, the Class~I/Class~II number ratio
     in our region is similar to that of the Serpens Main core, found to be among the highest in a
     survey of 60 cluster cores \citep{gutermuth2009}, and for which it has been suggested
     that star formation has proceeded in several phases \citep{casali1993,kaas2004}; the
     clustered protostar population having  formed in a recent burst.
     
     The apparently most active submm core has a surface density of YSOs $\sim$
     100~pc$^{-2}$, which is only slightly lower than that of both the NGC~1333 and the Serpens
     Main cores of $\sim$~140~pc$^{-2}$ \citep[][adjusted to updated distances]{gutermuth2009}.
     By assuming an average stellar mass of 1~M$_{\odot}$ we estimate the current star formation
     efficiency (SFE) locally in this core to be about 6\%.
     For comparison a SFE of 13-15\% was found for the NGC~1333 cores \citep{jorgensen2008}.

     The large number of extended molecular hydrogen objects we find in  G120.69+2.66
     reminds us of those in NGC~1333 and Serpens Main, i.e. well studied outflows with protostellar
     jets from Class~0/I sources \citep{davis2008,green2024}. We can only speculate about the
     origin of some of the extended molecular hydrogen objects in G120.69+2.66 that do not have
     an obvious driving source, whether they are protostellar jets, and also to what extent there
     are embedded protostars that have escaped our detection and classification.
     
     With the limited observations currently available in this study, we are most likely just ``scraping
     the surface'' of the YSO population. In spite of that, we find striking similarities with the youngest
     nearby embedded clusters and envision a similar star formation scenario. Deep mid- and far-IR
     observations are required to further investigate this.

\section{Summary}
\label{summary}

   \begin{enumerate}
   \item The IRAS~00267+6511 region, located towards the outer Galaxy at a distance of 1.1 kpc,
     comprises embedded on-going star formation. Both millimetre
     (CO and CS) lines and 850$\mu$m continuum emission reveal the presence of dense gas and
     dust in a molecular cloud containing one dense clump (PGCC G120.67+2.66) resolved into
     five SCUBA-2 cores.
     
   \item Near-- and mid-IR photometry is used to identify 38 YSOs with infrared excesses
     and/or variability, giving an average stellar surface density of 24~YSO~pc$^{-2}$ and
     reaching up to 113~YSO~pc$^{-2}$ in the richest populated submm core (\#2),
      for which we estimate the star formation efficiency to be $\sim$ 6\%. 
     
   \item  We classify 9 bona-fide Class~Is and 8 bona-fide Class~IIs, and most of the remaining
     YSOs are likely to be of Class~II category.
     The high Class~I/Class~II number ratio, as well as the spatial distribution of the YSOs
     with respect to the dense submm cloud cores, suggests that star formation is still in a very
     early stage. The region is likely younger than 1 Myr.

      \item Narrow-band H$_2$ line imaging at 2.122~$\mu$m detects a number
       of features, jet-like structures as well as isolated knots. The morphology and proximity to
       protostars is used to interpret flow structure and possible originating YSOs. The number of
       flows suggest recent episodes of mass accretion in several of the YSOs in the region. 

   \item Near-IR imaging reveals in one of the dense cores an hour-glass shaped nebula, likely
     the result of a deeply embedded YSO carving out its dust cocoon through a bipolar outflow.
      The H$_2$ images present a collimated jet emerging from inside the cone.
      
      \item G120.69+2.66 appears to belong to the group of more active Planck cold clumps that are currently sites with on-going star formation in the youngest protostellar stages. 
 
   \item We have discovered a very bright and reddened object seen towards the centre of this
     star formation region. Its spectrum shows it to have a cool temperature, we assume the
     infrared spectral type is around M2, and it is likely a red supergiant. If so, it is located
     far behind the cloud in the outer Galaxy and was previously unknown, despite its brightness,
     due to the very high foreground extinction in this region.
   \end{enumerate}

\begin{acknowledgements}
We thank two anonymous referees for thoughtful remarks.

  Part of this work was supported by Fundação para a Ciência e a Tecnologia (FCT) through the
  research grants UIDB/04434/2020 and UIDP/04434/2020. JLY acknowledges support from Onsala Space
  Observatory for the provisioning of its facilities/observational support. The Onsala Space
  Observatory national research infrastrcuture is funded through Swedish Research Council grant
  No 2017-00648.

  Partly based on observations made with the Nordic Optical Telescope, owned in
  collaboration by the University of Turku and Aarhus University, and operated
  jointly by Aarhus University, the University of Turku and the University of Oslo,
  representing Denmark, Finland and Norway, the University of Iceland and Stockholm
  University at the Observatorio del Roque de los Muchachos, La Palma, Spain, of
  the Instituto de Astrofisica de Canarias. We thank the NOT students Niilo Koivisto and Benjamin Nobre Hauptmann for contributing
to the NOTCam observations.
  
  This publication makes use of data products from the Wide-field Infrared Survey Explorer,
  which is a joint project of the University of California, Los Angeles, and the Jet
  Propulsion Laboratory/California Institute of Technology, funded by the National Aeronautics
  and Space Administration.

  This publication makes use of data products from the Two Micron All Sky Survey, which is a
  joint project of the University of Massachusetts and the Infrared Processing and Analysis
  Center/California Institute of Technology, funded by the National Aeronautics and Space
  Administration and the National Science Foundation.

  This research has made use of the SIMBAD database, operated at CDS, Strasbourg, France.

  This research has made use of the VizieR catalogue access tool, CDS,
  Strasbourg, France (DOI : 10.26093/cds/vizier). The original description 
  of the VizieR service was published in 2000, A\&AS 143, 23.
  
  This work has made use of data from the European Space Agency (ESA) mission
  {\it Gaia} (\url{https://www.cosmos.esa.int/gaia}), processed by the {\it Gaia}
  Data Processing and Analysis Consortium (DPAC,
  \url{https://www.cosmos.esa.int/web/gaia/dpac/consortium}). Funding for the DPAC
  has been provided by national institutions, in particular the institutions
  participating in the {\it Gaia} Multilateral Agreement.
  
\end{acknowledgements}

\bibliographystyle{aa}

\bibliography{export-bibtex-00267} 

\begin{table*}
  \caption{The 17 YSOs found from WISE mid-IR colours, listed together with 2MASS near-IR
    colours and errors when available. The YSO class is given in the last column. See
    Sect.~\ref{mir-results} for details.}
  \label{wise-ysos}
  {\small
  
  \begin{tabular}{lrrrrrrrrrrr}
  \hline
  \hline
\colzero WISE ID\colb  $J-H$\colc $\sigma_{J-H}$\cold $H-K_S$\cole  $\sigma_{H-K_S}$\colf [3.4] - [4.6]\colg $\sigma_{3.4-4.6}$\colh [4.6] - [12]\coli $\sigma_{4.6-12}$\colj [4.6]\colk $\sigma_{4.6}$\coll Class\eol
\colzero        \colb  mag\colc mag\cold mag\cole mag\colf mag\colg mag\colh mag\coli mag\colj mag\colk mag\coll \eol
\extline
\hline
\colzero     J002941.83+652631.9$^{{c}}$\colb  1.16\colc  0.09\cold  0.78\cole  0.07\colf  1.23\colg  0.03\colh  2.67\coli  0.05\colj  11.60\colk  0.02\coll I$^{{a}}$\eol
\colzero     J002954.81+652659.2\colb      \colc      \cold      \cole      \colf  1.92\colg  0.08\colh  3.72\coli  0.04\colj  12.26\colk  0.03\coll  I\eol
\colzero     J002952.30+652725.3\colb  2.44\colc      \cold  1.44\cole  0.06\colf  0.88\colg  0.03\colh  3.29\coli  0.03\colj  11.39\colk  0.02\coll  I\eol
\colzero     J002950.58+652719.6$^{{c}}$\colb  2.01\colc  0.15\cold  2.30\cole  0.08\colf  1.89\colg  0.03\colh  2.90\coli  0.02\colj  8.42\colk  0.02\coll   I\eol
\colzero     J002932.58+652510.0\colb      \colc      \cold      \cole      \colf  2.81\colg  0.03\colh  3.76\coli  0.02\colj  9.16\colk  0.02\coll   I\eol
\colzero     J002929.28+652746.0\colb  2.57\colc      \cold  1.84\cole      \colf  2.56\colg  0.03\colh  3.51\coli  0.02\colj  8.33\colk  0.02\coll   I\eol
\colzero     J002931.45+652718.4\colb  1.67\colc      \cold  1.82\cole  0.10\colf  1.40\colg  0.03\colh  2.81\coli  0.02\colj  8.75\colk  0.02\coll   I\eol
\colzero     J002937.06+652813.1\colb  1.88\colc      \cold  1.71\cole      \colf  1.62\colg  0.03\colh  2.12\coli  0.02\colj  8.18\colk  0.02\coll   I$^{{b}}$\eol
\colzero     J002940.09+652811.1$^{{c}}$\colb  1.06\colc  0.12\cold  0.94\cole  0.09\colf  1.27\colg  0.03\colh  2.62\coli  0.03\colj 10.73\colk  0.02\coll   I\eol
\colzero     J002949.31+652528.5\colb  1.93\colc      \cold  1.11\cole  0.05\colf  0.76\colg  0.03\colh  1.86\coli  0.03\colj  9.65\colk  0.02\coll   II\eol
\colzero     J002952.88+652539.9\colb  1.65\colc      \cold  1.41\cole  0.10\colf  0.96\colg  0.03\colh  1.94\coli  0.05\colj 11.63\colk  0.02\coll   II\eol
\colzero     J002951.92+652702.7$^{{c}}$\colb  1.27\colc  0.07\cold  0.57\cole  0.06\colf  0.79\colg  0.03\colh  2.32\coli  0.05\colj 11.74\colk  0.02\coll   II\eol
\colzero     J002932.47+652637.7$^{{c}}$\colb  0.91\colc  0.04\cold  0.56\cole  0.04\colf  0.65\colg  0.03\colh  1.80\coli  0.03\colj  9.07\colk  0.02\coll   II\eol
\colzero     J002925.19+652637.7\colb  1.08\colc  0.06\cold  0.81\cole  0.05\colf  0.70\colg  0.03\colh  1.72\coli  0.06\colj 11.57\colk  0.02\coll   II\eol
\colzero     J002928.25+652634.1$^{{c}}$\colb  1.31\colc  0.05\cold  0.70\cole  0.05\colf  0.57\colg  0.04\colh  2.00\coli  0.05\colj 11.39\colk  0.02\coll   II\eol
\colzero     J002937.69+652621.7$^{{c}}$\colb  1.15\colc  0.05\cold  0.76\cole  0.05\colf  0.84\colg  0.03\colh  2.42\coli  0.03\colj 10.66\colk  0.02\coll   II\eol
\colzero     J002941.52+652730.9$^{{c}}$\colb  1.97\colc      \cold  1.29\cole      \colf  1.04\colg  0.03\colh  1.69\coli  0.03\colj  9.41\colk  0.02\coll   II\eol

\hline
\end{tabular}
{\footnotesize  Notes:}

\footnotetext{Has a deviating proper motion and is located outside the dense cores.}
\footnotetext{Candidate WISE originating source for the hour-glass shaped nebula, in which case both coordinates and 2MASS cross-correlation, here taken from the WISE catalogue, are uncertain, See Sect.~\ref{extended-sources}.} 
\footnotetext{YSO candidate in \citet{marton2016} and/or \citet{marton2019}.}
  }
\end{table*}

\begin{table*}
  \caption{The 21 additional YSO candidates found in the NOTCam and 2MASS sample, selected from
    either near-IR excess or variability or both. The cross-correlated 2MASS point source detection is
    given as an ID while the RA/DEC coordinates, as well as the near-IR magnitudes and colours
    are from the NOTCam data.}
  \label{notcam-ysos}
  {\small
  
  \begin{tabular}{lrrrrrrrrr}
  \hline
  \hline

  \cola 2MASS-ID\colb RAJ2000\colc DEJ2000\cold $H$\cole $\sigma_{H}$\colf $J-H$\colg $\sigma_{J-H}$\colh $H-K_S$\coli $\sigma_{H-K}$\colj YSO criterion\eol
\cola \colb deg\colc deg\cold mag\cole mag\colf mag\colg mag\colh mag\coli mag\colj \eol
\hline
\cola 00292061+6528458\colb    7.33595\colc   65.47936\cold 15.85\cole  0.03\colf  1.18\colg  0.05\colh  0.92\coli  0.16\colj  NIR excess\eol
\cola 00292316+6526123$^{{a}}$\colb    7.34653\colc   65.43672\cold 14.83\cole  0.03\colf  1.22\colg  0.04\colh  0.58\coli  0.10\colj  Variable $H$ \eol
\cola 00292601+6526014\colb    7.35837\colc   65.43375\cold 12.12\cole  0.03\colf  0.54\colg  0.03\colh  0.43\coli  0.06\colj  NIR excess\eol
\cola 00292700+6526051\colb    7.36250\colc   65.43478\cold 14.53\cole  0.03\colf  1.03\colg  0.04\colh  0.61\coli  0.09\colj  NIR excess\eol
\cola 00292916+6526190\colb    7.37149\colc   65.43862\cold 14.23\cole  0.03\colf  0.76\colg  0.04\colh  0.55\coli  0.08\colj  NIR excess\eol
\cola 00292956+6526017\colb    7.37326\colc   65.43379\cold 15.19\cole  0.03\colf  1.59\colg  0.04\colh  0.86\coli  0.10\colj  Variable $H$ \eol
\cola 00293083+6526490\colb    7.37846\colc   65.44699\cold 12.72\cole  0.03\colf  2.31\colg  0.04\colh  1.30\coli  0.06\colj  NIR excess \eol
\cola 00293280+6527210\colb    7.38667\colc   65.45583\cold 15.25\cole  0.03\colf  2.26\colg  0.05\colh  1.02\coli  0.10\colj  Variable $K_S$ \eol
\cola                 \colb    7.38748\colc   65.47388\cold 18.80\cole  0.06\colf  0.73\colg  0.08\colh  0.70\coli  0.09\colj  NIR excess\eol
\cola 00293391+6527157\colb    7.39133\colc   65.45448\cold 16.59\cole  0.04\colf  2.66\colg  0.08\colh  1.80\coli  0.06\colj  NIR excess\eol
\cola                 \colb    7.39428\colc   65.46428\cold 16.55\cole  0.04\colf  1.46\colg  0.06\colh  1.00\coli  0.06\colj  NIR excess\eol
\cola 00293683+6528128\colb    7.40348\colc   65.47021\cold 15.91\cole  0.04\colf  2.82\colg  0.09\colh  1.95\coli  0.10\colj  NIR excess\eol
\cola 00293773+6528134\colb    7.40753\colc   65.47038\cold 15.37\cole  0.04\colf  1.96\colg  0.06\colh  1.32\coli  0.06\colj  NIR excess\eol
\cola 00293889+6528160\colb    7.41210\colc   65.47108\cold 17.05\cole  0.04\colf  2.01\colg  0.07\colh  1.50\coli  0.06\colj  NIR excess\eol
\cola 00293975+6525030\colb    7.41569\colc   65.41750\cold 14.82\cole  0.03\colf  0.81\colg  0.04\colh  0.03\coli  0.15\colj  Variable $J$ + $K_S$\eol
\cola                 \colb    7.41967\colc   65.46547\cold 18.58\cole  0.05\colf  0.63\colg  0.07\colh  0.51\coli  0.09\colj  NIR excess \eol
\cola 00294167+6527049\colb    7.42367\colc   65.45137\cold 14.61\cole  0.03\colf  1.23\colg  0.04\colh  0.23\coli  0.11\colj  Variable $H$ + $K_S$\eol
%
%
\cola 00294396+6526034\colb    7.43321\colc   65.43431\cold 13.95\cole  0.03\colf  0.97\colg  0.04\colh  0.60\coli  0.08\colj  NIR excess, bipolar H$_2$ jet\eol
\cola 00294693+6528014\colb    7.44546\colc   65.46706\cold 14.58\cole  0.03\colf  1.03\colg  0.04\colh -0.18\coli  0.13\colj  Variable $K_S$\eol
\cola 00294889+6525251\colb    7.45369\colc   65.42361\cold 15.88\cole  0.03\colf  2.11\colg  0.06\colh  1.32\coli  0.12\colj  NIR excess, Variable $H$ + $K_S$\eol
\cola 00295286+6526412\colb    7.47025\colc   65.44478\cold 13.43\cole  0.03\colf  1.96\colg  0.04\colh  0.80\coli  0.07\colj  Variable $K_S$\eol
\hline
\end{tabular}
   {\footnotesize  Notes:}
\footnotetext{YSO candidate in \citet{marton2019}.}
   }
\end{table*}

\end{document}